\documentclass[pra, twocolumn,preprintnumbers,amsmath,amssymb,showpacs,aps,pra,superscriptaddress]{revtex4}
\usepackage{mathrsfs}
\usepackage{mathrsfs}
\usepackage{braket}
\usepackage{graphicx}
\usepackage{dcolumn}
\usepackage{amsmath}
\usepackage{epsfig}
\usepackage{subfigure}
\usepackage{color}
\usepackage{epstopdf}
\usepackage{framed}

\newcommand {\abold}{\ensuremath \hat{\boldsymbol{a}}}

\begin{document}

\date{\today}

\title{Floodlight quantum key distribution:  A practical route to Gbps secret-key rates}
\author{Quntao Zhuang}
\email{quntao@mit.edu}
\affiliation{Research Laboratory of Electronics, Massachusetts Institute of Technology,
77 Massachusetts Avenue, Cambridge, Massachusetts 02139, USA}
\affiliation{Department of Physics, Massachusetts Institute of Technology, Cambridge, Massachusetts 02139, USA}
\author{Zheshen Zhang}
\affiliation{Research Laboratory of Electronics, Massachusetts Institute of Technology,
77 Massachusetts Avenue, Cambridge, Massachusetts 02139, USA}
\author{Justin Dove}
\affiliation{Research Laboratory of Electronics, Massachusetts Institute of Technology,
77 Massachusetts Avenue, Cambridge, Massachusetts 02139, USA}
\author{Franco N. C. Wong}
\affiliation{Research Laboratory of Electronics, Massachusetts Institute of Technology,
77 Massachusetts Avenue, Cambridge, Massachusetts 02139, USA}
\author{Jeffrey H. Shapiro}
\affiliation{Research Laboratory of Electronics, Massachusetts Institute of Technology,
77 Massachusetts Avenue, Cambridge, Massachusetts 02139, USA}

\begin{abstract} 
The channel loss incurred in long-distance transmission places a significant burden on quantum key distribution (QKD) systems:  they must defeat a passive eavesdropper who detects all the light lost in the quantum channel and does so without disturbing the light that reaches the intended destination.   The current QKD implementation with the highest long-distance secret-key rate meets this challenge by transmitting no more than one photon per bit [Opt.~Express {\bf 21,} 24550--24565 (2013)].  As a result, it cannot achieve the Gbps secret-key rate needed for one-time pad encryption of large data files unless an impractically large amount of multiplexing is employed. We introduce floodlight QKD (FL-QKD), which floods the quantum channel with a high number of photons per bit distributed over a much greater number of optical modes. FL-QKD offers security against the optimum frequency-domain collective attack by transmitting less than one photon per mode and using photon-coincidence channel monitoring, and it is completely immune to passive eavesdropping.  More importantly, FL-QKD is capable of a 2\,Gbps secret-key rate over a 50\,km fiber link, \emph{without} any multiplexing, using available equipment, i.e., no new technology need be developed.  FL-QKD achieves this extraordinary secret-key rate by virtue of its unprecedented secret-key efficiency, in bits per channel use, which exceeds those of state-of-the-art systems by two orders of magnitude.  
\end{abstract}

\pacs{ 03.67.Hk, 03.67.Dd, 42.50.Lc}

\maketitle

\section{Introduction}
One-time pad (OTP) encryption provides information-theoretically secure message transmission \cite{Shannon1949}, but key distribution is its Achilles' heel.  Quantum key distribution (QKD) permits remote parties (Alice and Bob) to share a random bit string---the key needed for OTP encryption---with security vouchsafed by quantum mechanics \cite{Bennett1984,Ekert1991,Gisin2002,Grosshans2002}.  Unfortunately, the demonstrated secret-key rates of long-distance QKD systems fall far short of the Gbps rates needed for OTP encryption of large data files, as seen from the following state-of-the-art achievements.  In discrete-variable QKD (DV-QKD), the best result to date is Lucamarini \emph{et al.}'s decoy state Bennett-Brassard 1984 (BB84) system, which used a 1\,Gbps source rate but only realized a 1\,Mbps secret-key rate over a 50-km-long fiber \cite{Lucamarini2013}.  In continuous-variable QKD (CV-QKD), the best result to date is from Huang \emph{et al.}, who reported a 1\,Mbps secret-key rate at 25\,km path length using a 50\,Mbaud source rate \cite{Huang2015}, with 90\,kbps expected at 50\,km in the asymptotic (infinite block-length) regime.  

Focusing, for the moment, on DV-QKD systems---owing to their greater demonstrated capability over long distances---it is easy to identify why Gbps rates are beyond their state-of-the-art grasp:  they transmit no more than $\sim$1\,photon/bit.  One justification for this self-imposed limit is that these systems must defeat the undetectable passive eavesdropper.  QKD security analyses afford the eavesdropper (Eve) all things consistent with the laws of physics. In particular, a passive Eve could replace the transmissivity $\kappa \ll 1$ optical fiber connecting Alice and Bob with a lossless long-distance coupler that allows her to capture and measure a fraction $1-\kappa$ of Alice's transmitted light while routing the remaining fraction $\kappa$ to Bob without disturbance.  With no disturbance of the light that Bob receives, Eve does not create the telltale errors that reveal her eavesdropping.  In principle, such a coupler could be constructed to mimic---insofar as Alice and Bob are concerned---the propagation characteristics of the fiber that it replaced.  Thus Alice and Bob could not detect Eve's presence via channel monitoring, e.g., with an optical time-domain reflectometer.  So, were Alice to ignore the potential presence of the undetectable passive eavesdropper and make a many-photons-per-bit BB84 transmission to Bob through this lossy quantum channel, then Eve could easily obtain a near-perfect measurement of \emph{all} of Alice's bits.  

We regard secret-key rate, in bits per second, as QKD systems' preeminent figure of merit:  unless Gbps rates over metropolitan-area spans can be realized, OTP-encrypted transmission of large data files will not reach widespread usage.  Existing QKD systems operating over long--distance connections \emph{might} be pushed to Gbps secret-key rates, but doing so would require impractically large amounts of wavelength-division multiplexing (WDM).  Consider scaling Lucamarini \emph{et al}.'s BB84 system \cite{Lucamarini2013} to a 10\,Gbps source rate achieving a 10\,Mbps secret-key rate over a 50\,km fiber link.  That system would require 100 WDM channels to yield a 1\,Gbps secret-key rate---while 1000 such channels would be needed at the original source rate---each with its own single-photon detection setup.  A similar scaling of Huang \emph{et al}.'s CV-QKD system \cite{Huang2015}---to a 10\,Gbaud source rate that achieves 18\,Mbps secret-key rate over a 50\,km fiber link in the asymptotic regime---implies that more than 50 WDM channels would be needed to obtain a 1\,Gbps secret-key rate.  

In this paper we introduce floodlight quantum key distribution (FL-QKD), and show that it offers a practical route to Gbps secret-key rates over metropolitan-area distances with security against the optimum frequency-domain collective attack and without the need for multiplexing.   How does FL-QKD realize this extraordinary secret-key rate?  It derives from FL-QKD's secret-key efficiency, in bits per channel use, being two order of magnitude higher than those of state-of-the-art systems.  In particular, FL-QKD floods the Alice-to-Bob channel with broadband light---whose bandwidth is much greater than the modulation rate---containing many photons per bit.   Its immunity to the undetectable passive-eavesdropping attack then comes from that high number of transmitted photons per bit being distributed over a much greater number of optical modes to make that transmission have low brightness, i.e., less than one photon per mode.  FL-QKD also employs photon-coincidence channel monitoring on the Alice-to-Bob channel, to ensure security against the active component of a frequency-domain collective attack, in which Eve can inject her own light into Bob's terminal and tries to obtain his bit string from the modulated version of that light which is contained in what she taps from the Bob-to-Alice channel.  More importantly, we show that FL-QKD can support a 2\,Gbps secret-key rate over a 50-km-long fiber link against the optimum frequency-domain collective attack, and that it can be implemented with available equipment, i.e., no new technology need be developed.  In short, FL-QKD opens the possibility for OTP encryption of large data files for secure transmission over metropolitan-area distances at Gbps rates.  

The remainder of the paper is organized as follows.  Sections~II through V present, in succession, a description of the FL-QKD protocol, its security analysis, its secret-key rate behavior, and some concluding discussion.  For the sake of readability, we have relegated all detailed analysis to a series of appendices.

\section{Protocol Description}  Figure~\ref{setup} shows FL-QKD's quantum channel setup in the presence of a frequency-domain collective attack.  Alice and Bob use this setup to generate their raw key and to bound Eve's Holevo information.  Not shown in this figure is the tamper-proof classical channel that Alice and Bob use for reconciliation.  Neither that procedure nor FL-QKD's subsequent privacy amplification step will be described herein, because they are merely higher rate versions of standard practice in QKD.  

\begin{figure}
\includegraphics[width=3.25in]{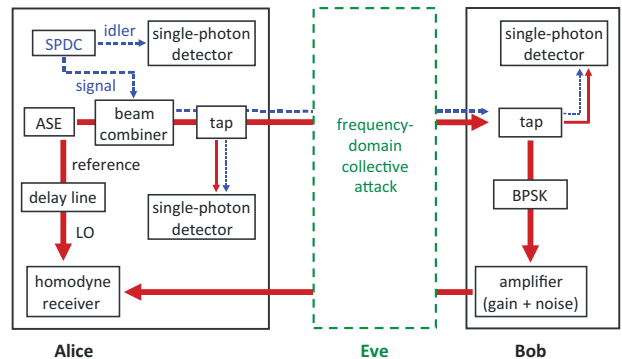}
\caption{(color online). \label{setup}Quantum channel setup for FL-QKD under frequency-domain collective attack.  ASE:  amplified spontaneous emission source.  SPDC:  spontaneous parametric downconverter.  BPSK:  binary phase-shift keying.  LO:  local oscillator.}
\end{figure}

Raw key generation in FL-QKD occurs as follows.  Alice sends unmodulated, continuous-wave (cw) light over optical fiber to Bob, who imposes a random bit string on that light by means of binary phase-shift keying (BPSK) at $R$ bps, amplifies the modulated light (to overcome return-path loss), and returns it to Alice over optical fiber.  FL-QKD's security against a frequency-domain collective attack, and its high secret-key rate, come from the composite nature of Alice's source plus the data that Alice and Bob obtain from their channel monitors, which are used to ensure the integrity of the Alice-to-Bob channel, i.e, the near-perfect correlation between the light reaching Bob and the reference retained by Alice.  So, to complete our protocol description, we will characterize Alice's source and Alice and Bob's channel monitors.

Alice uses an optical amplifier to produce a high-brightness ($\gg 1$ photon s$^{-1}$\,Hz$^{-1}$) single spatial-mode beam of amplified spontaneous emission (ASE) noise with a $W$-Hz-bandwidth flat spectrum.  She uses a cw spontaneous parametric downconverter (SPDC) to produce quadrature-entangled, single spatial-mode signal and idler beams that have bandwidth $W$ flat spectra, with the former having the same center frequency as her ASE source.  Alice directs the idler beam to a single-photon detector that is part of her channel monitor.  She uses a beam combiner to merge a low-brightness ($\ll 1$ photon s$^{-1}$\,Hz$^{-1}$) portion of her ASE light with her SPDC's signal light resulting in an $n$:1 ASE-to-SPDC-ratio output with $n\gg 1$.  She sends a small fraction of her combined ASE-SPDC light to another single-photon detector (also part of her channel monitor), and transmits the remaining portion of her ASE-SPDC light to Bob.  Alice stores the high-brightness portion of her initial ASE light in an optical delay-line fiber (whose delay matches that of the Alice-to-Bob-to-Alice roundtrip) for use as the local oscillator (LO) in a broadband homodyne receiver.  She employs optical amplification, as needed, so that her LO retains its high-brightness character without appreciable degradation, see App.~A.3 for details.  Prior to BPSK modulation, Bob routes a small fraction of the light he receives to the single-photon detector that is his channel monitor.  

Alice and Bob use their channel monitors to measure the singles rates, $S_I$ for Alice's idler beam, $S_A$ for Alice's tap on her transmitted beam, and $S_B$ for Bob's tap on his received beam.  They also use their monitors to obtain $C_{IA}$ and $\widetilde{C}_{IA}$, the time-aligned and time-shifted coincidence rates between Alice's idler and the tap on her transmitted beam, and $C_{IB}$ and $\widetilde{C}_{IB}$, the time-aligned and time-shifted coincidence rates between Alice's idler and Bob's tap on his received beam, in both cases employing a $T_g$-duration coincidence gate and accounting for the relevant propagation delays in the appropriate manners.  From these rates they compute
\begin{equation}
f_E = 1-\frac{[C_{IB}-\widetilde{C}_{IB}]/S_B}{[C_{IA}-\widetilde{C}_{IA}]/S_A},
\end{equation}
which will be shown below to quantify the integrity of the Alice-to-Bob channel.  

\section{Security Analysis} 
\begin{figure}
\includegraphics[width=3in]{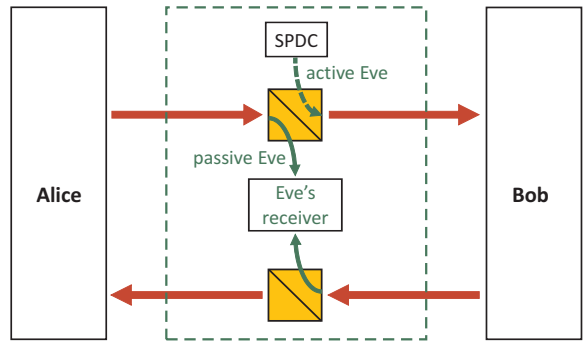}
\caption{(color online).  \label{realization}Realization of Eve's optimum frequency-domain collective attack.  SPDC:  spontaneous parametric downconverter. Eve's SPDC signal beam (shown) is coupled to Bob through a beam splitter, while her SPDC idler beam (not shown) is retained for use in her receiver.}
\end{figure}

As detailed in App.~B, Eve's general frequency-domain collective attack is as follows.  Eve first establishes lossless connections between her equipment and the communicating parties in both the forward (Alice-to-Bob) and backward (Bob-to-Alice) channels.  In the forward path, she performs a general unitary transformation that, during each of Bob's bit intervals, acts in an independent, identically distributed manner on the $M=W/R$ frequency modes of Alice's transmitted light.   In particular, the inputs to that unitary transformation are Alice's transmitted field and Eve's $K$ vacuum-state ancilla fields.  Eve retains the $K$ ancilla fields that emerge from this unitary operation and sends the remaining field to Bob.  She completes her attack with a collective measurement on her stored ancilla fields and the light she taps from the Bob-to-Alice channel.  Here we note, see App.~E, that $f_E$ is an intrusion parameter that quantifies Eve's degradation of the phase-sensitive cross-covariance between Alice's idler and Bob's received light from what it would be were Eve only mounting a passive attack.
Furthermore, we show in App.~C.2 that Eve's optimum frequency-domain collective attack---one that maximizes her Holevo information for a given photon flux and $f_E$ value---is in fact Gaussian and can be realized by her using an SPDC source, injecting its signal light into Bob through a beam splitter in the Alice-to-Bob fiber, while retaining her idler for a collective measurement with the light she taps from the Alice-to-Bob and Bob-to-Alice fibers, see Fig.~\ref{realization}. For this optimum attack, $f_E$ equals Eve's injection fraction, viz., the fraction of light entering Bob's terminal that is due to her \cite{footnote1}.   Hence that configuration will be employed throughout the security analysis below.  (Interestingly, this SPDC beam-splitter attack has the same structure as the entangling-cloner attack on CV-QKD \cite{Navascues2005}.)

We will be concerned with optimized performance for Alice and Bob against Eve's optimum frequency-domain collective attack without regard for finite-key effects.  (For FL-QKD's $\sim$Gbps secret-key rates, finite-key effects become inconsequential for key-generation sessions as short as a few seconds.) Thus, following standard practice for assessing security against collective attacks (see, e.g., \cite{Zhang2014,Pirandola2015a}), we will find $\Delta I_{AB}^{\rm LB}$, a lower bound on Alice and Bob's secret-key rate, from 
\begin{equation}
\Delta I_{AB}^{\rm LB} = \beta I_{AB} - \chi_{EB}^{\rm UB},
\label{DeltaIab}
\end{equation}
where $I_{AB}$ is Alice and Bob's Shannon-information rate, $\beta$ is Alice and Bob's reconciliation efficiency, and $\chi_{EB}^{\rm UB}$ is an upper bound on Eve's Holevo-information rate for her optimum frequency-domain collective attack.  Before doing so, let us provide some simple intuition about how FL-QKD can be secure against individual passive or active attacks.  

We will limit our consideration of these individual attacks to low-brightness operation (the ASE-SPDC light Alice sends to Bob has $N_S \ll 1$ photon s$^{-1}$\,Hz$^{-1}$) in a lossy scenario (channel transmissivity $\kappa_S \ll 1$) with Alice's source bandwidth $W$ greatly exceeding Bob's BPSK modulation rate $R$.  For Eve's passive attack, we neglect the small amount of SPDC light in Alice's transmission and the small amounts tapped by Alice and Bob for their channel monitors.  Alice's homodyne receiver and Eve's optimum quantum receiver then have error probabilities satisfying $\Pr(e)_{\rm Alice}^{\rm hom} \sim \exp(-W\kappa_SN_SG_B/RN_B)/2$ \cite{AppD} and $\Pr(e)_{\rm Eve}^{\rm pass} \sim \exp(-4W\kappa_SN_S^2G_B/RN_B)/2$ \cite{Shapiro2009}, where $G_B\gg 1$ and $N_B \ge G_B-1$ are the gain and ASE output-noise brightness of Bob's optical amplifier.  Because $\ln[\Pr(e)_{\rm Alice}^{\rm hom}]/\ln[\Pr(e)_{\rm Eve}^{\rm pass}] \sim 1/4N_S$, we see that low-brightness ($N_S \ll1$) operation affords Alice and Bob a considerable advantage over Eve.  Physically, this advantage is due to the $N_S \ll 1$ low-brightness condition's making Eve unable to obtain a high-brightness reference---from the light she taps from the Alice-to-Bob fiber---with which to detect Bob's BPSK modulation.  Later, we will see that this low-brightness condition ensures that Eve's Holevo information rate for her undetectable passive-eavesdropping attack falls far below Alice and Bob's Shannon information rate.  In other words, as claimed earlier, FL-QKD's transmitting less than one photon per mode makes it immune to the attack that has driven the highest-rate, long-distance QKD system to limit its transmissions to $\sim$1\,photon/bit.

For Eve's active attack, we employ the conditions applied above and, in addition, presume that Alice and Bob's channel monitors constrain their adversary's light injection to a small fraction, $f_E \ll1$, of the light entering Bob's terminal.   The error probability of Alice's homodyne receiver will then obey $\Pr(e)_{\rm Alice}^{\rm hom} \sim \exp(-W(1-f_E)\kappa_SN_SG_B/RN_B)/2$.  Eve's optimum quantum receiver---for an individual attack in the Fig.~\ref{realization} setup using her optimum SPDC-injection strategy in conjunction with a tap on just the Bob-to-Alice channel---then has error probability $\Pr(e)_{\rm Eve}^{\rm act} \sim \exp(-4Wf_E\kappa_SN_SG_B/RN_B)/2$.  Now we find that $\ln[\Pr(e)_{\rm Alice}^{\rm hom}]/\ln[\Pr(e)_{\rm Eve}^{\rm act}] \sim (1-f_E)/4f_E$, which is highly favorable to Alice and Bob when their channel monitors limit Eve to $f_E \ll 1$.  

Having provided some individual-attack insights into FL-QKD's security, we return to the task of assessing our protocol's security analysis when Eve mounts her optimum frequency-domain collective attack.  To evaluate Alice's error probability under that attack, we note the number of independent modes that contribute to the light Alice receives from Bob being much greater than one---for $W = 2\,$THz with $R \le 10\,$Gbps, as we will assume below, we get $M = W/R \ge 200$---justifies a central limit theorem argument that makes Alice's error probability satisfy \cite{Zhang2013}
\begin{equation}
\Pr(e)_{\rm Alice}^{\rm hom} = Q\!\left(\frac{\mu_0 - \mu_1}{\sigma_0 + \sigma_1}\right),
\end{equation}
where $\mu_b$ and $\sigma_b$ for $b=0,1$ are the means and standard deviations of Alice's homodyne measurement when Bob's bit values (phase modulations) are equally likely to be 0 (0\,rad phase shift) or 1 ($\pi$\,rad phase shift), and $Q(x) = \int_x^\infty\,{\rm d}t\,e^{-t^2/2}/\sqrt{2\pi}$.  See App.~D for the $\{\mu_b\}$ and $\{\sigma_b\}$ with all losses included.  With Alice's error probability in hand, Alice and Bob's Shannon-information rate is found from 
\begin{align}
I_{AB} &= R\!\left[1+\Pr(e)_{\rm Alice}^{\rm hom}\log_2(\Pr(e)_{\rm Alice}^{\rm hom})\right.\nonumber \\
& \left.+ (1-\Pr(e)_{\rm Alice}^{\rm hom})\log_2(1-\Pr(e)_{\rm Alice}^{\rm hom})\right].
\end{align}

Eve's Holevo-information rate about Bob's bit string for her optimum collective attack is 
\begin{equation}
\chi_{EB} =R\! \left[S({\boldsymbol \rho}_E) - \sum_{b=0}^1S({\boldsymbol \rho}^{(b)}_E)/2\right],
\end{equation}
where $S(\cdot)$ denotes von Neumann entropy.  Here, ${\boldsymbol \rho}_E^{(b)}$ is Eve's conditional joint density operator---when Bob transmits a single bit with value $b = 0$ or 1---for the $3M$ modes available to her that are associated with that bit, viz., $M$ modes each from her retained idler, the light she collects from the Alice-to-Bob fiber, and the light she collects from the Bob-to-Alice fiber. Her unconditional joint density operator for those $3M$ modes is then ${\boldsymbol \rho}_E = \sum_{b=0}^1{\boldsymbol \rho}^{(b)}_E/2$.  The ${\boldsymbol \rho}^{(b)}_E$ are zero-mean Gaussian states whose von Neumann entropies are easily found by symplectic diagonalization \cite{Pirandola2008}, as explained in App.~C. The unconditional state, ${\boldsymbol \rho}_E$ is zero mean but \emph{not} Gaussian, making its von Neumann entropy quite difficult to evaluate.  However, that state's covariance matrix is easily found \cite{AppC}, and we know that $S({\boldsymbol \rho}_E) \le S({\boldsymbol \rho}_{E}^{\rm Gauss})$, where ${\boldsymbol \rho}_{E}^{\rm Gauss}$ is a zero-mean Gaussian state with the same covariance matrix as ${\boldsymbol \rho}_E$.  We can find $S({\boldsymbol \rho}_{E}^{\rm Gauss})$ by another symplectic diagonalization and so obtain 
\begin{equation}
\chi_{EB} \le \chi_{EB}^{\rm UB} = R\min\!\left[S({\boldsymbol \rho}_{E}^{\rm Gauss}) - \sum_{b=0}^1S({\boldsymbol \rho}^{(b)}_E)/2,1\right],
\end{equation}
where we have used $S({\boldsymbol \rho}_E) - \sum_{b=0}^1S({\boldsymbol \rho}^{(b)}_E)/2 \le 1$, which follows from that term's being Eve's Holevo information about a single-bit transmission from Bob.

\section{Secret-Key Rates} 
\begin{figure}
\vspace*{.05in}
\includegraphics[width=1.815in]{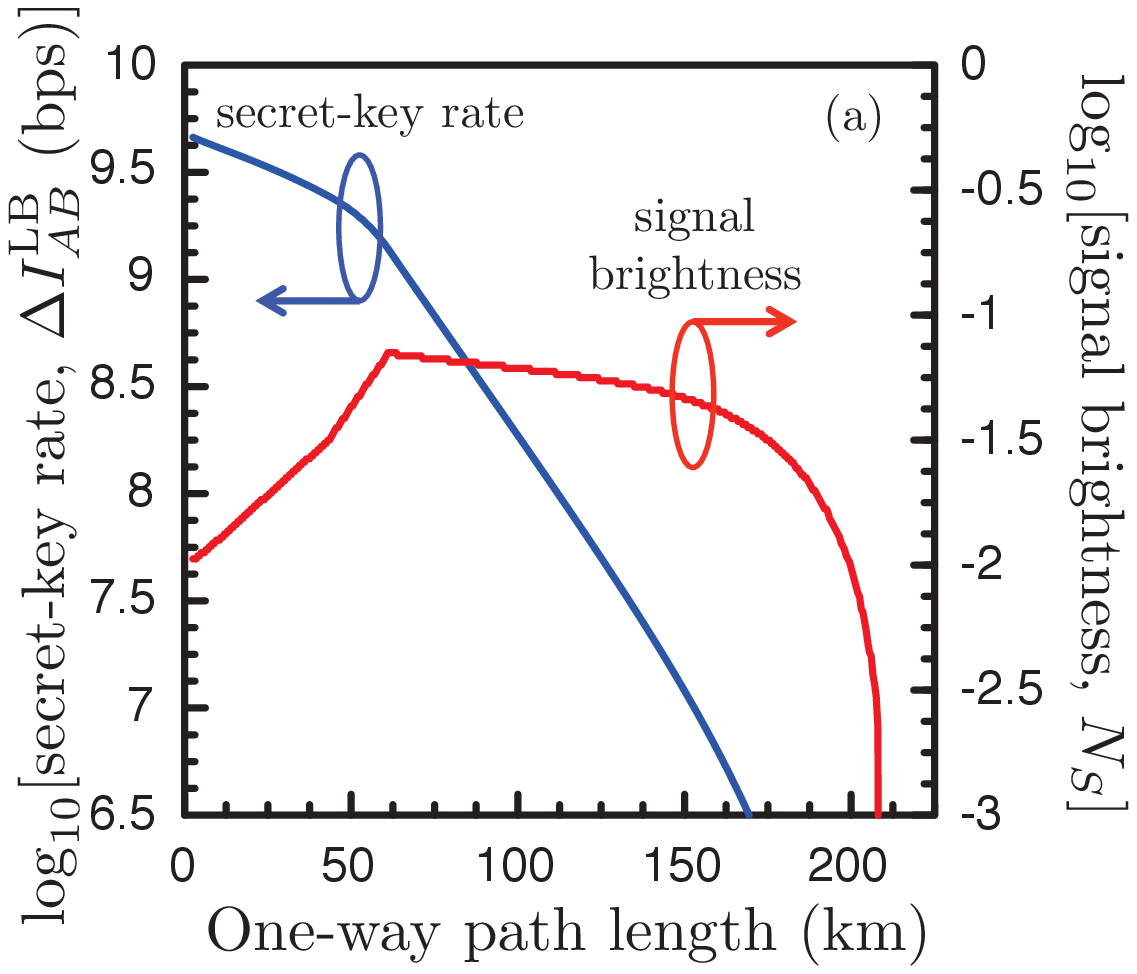}\hspace{.05in}\includegraphics[width=1.535in]{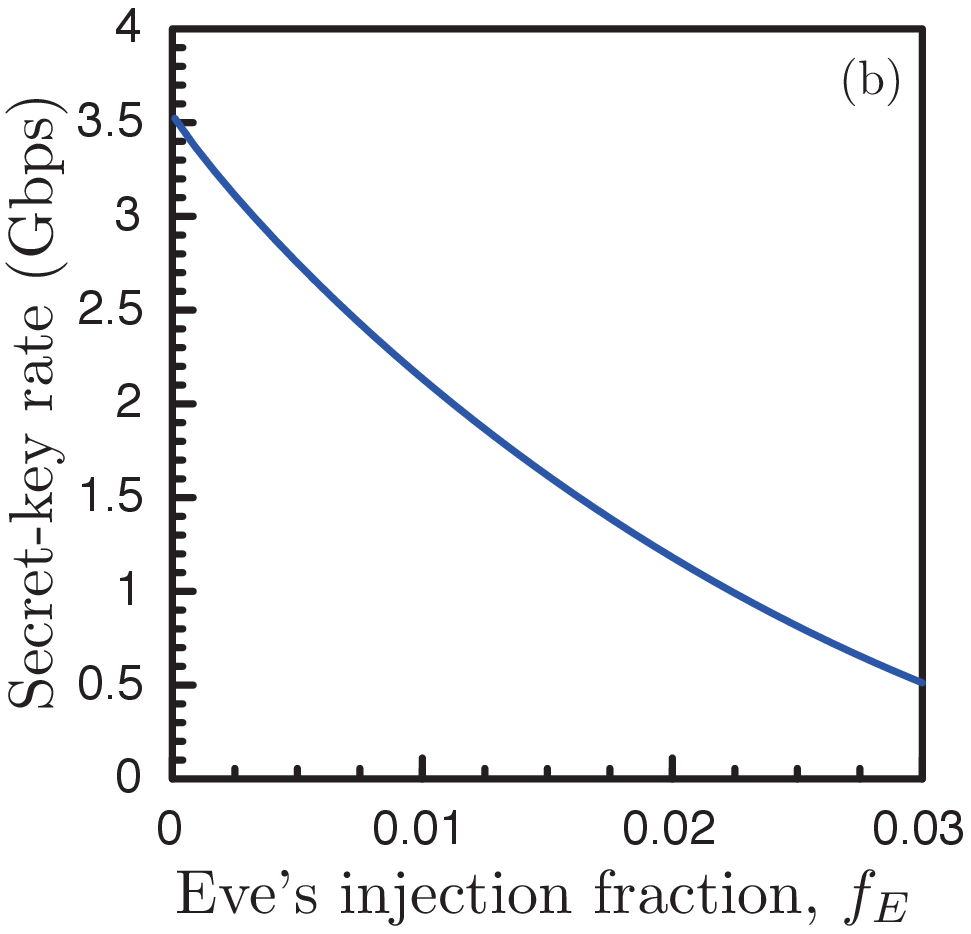}
\caption{(color online). \label{keyrates} (a) Lower bound on Alice and Bob's secret-key rate and Alice's optimum signal brightness when Eve mounts her optimum frequency-domain collective attack with injection fraction $f_E = 0.01$. (b) Lower bound on Alice and Bob's secret-key rate versus $f_E$ for a 50-km fiber link with all other parameters as in (a).}
\end{figure}

We are now ready to demonstrate the power of FL-QKD.  Figure~\ref{keyrates}(a) plots the lower bound from Eq.~(\ref{DeltaIab}) on Alice and Bob's secret-key rate versus one-way path length when Eve mounts her optimum collective attack, but Alice and Bob's channel monitoring ensures that Eve's injection fraction into Bob's terminal is $f_E=0.01$.  Also shown in that figure is a brightness versus path length plot for the light Alice sends to Bob.  These curves were obtained assuming that:  (1) Alice's ASE source and her SPDC signal light have flat spectra with the same center frequency and $W=2$\,THz bandwidth, and are combined in an $n$:1 ratio with $n=99$;  (2) the brightness of the light Alice sends to Bob and Bob's bit rate $R \le 10\,$Gbps are chosen to maximize their secret-key rate subject to the constraint that $\Pr(e)_{\rm Alice} \le 0.1$ to ensure the availability of a high-efficiency code for reconciliation \cite{Richardson2001}; (3) Bob's amplifier has $G_B = N_B = 10^4$; (4) Eve has replaced the $L$-km-long, 0.2\,dB/km fibers in the Alice-to-Bob and Bob-to-Alice channels with lossless fibers and $(1-f_E)\kappa_S$ and $\kappa_S$ transmissivity beam splitters, respectively, with $\kappa_S = 10^{-0.02L}$; (5) Alice taps 1\% of her combined ASE-SPDC light, and Bob taps 1\% of his received light, for channel monitoring; (6) Alice's homodyne receiver has an undegraded local oscillator with brightness $N_{\rm LO} = 10^4$ and efficiency 0.9; (7) $\beta=0.94$; and (8) the system is otherwise ideal.  

\begin{figure}
\includegraphics[width=1.69in]{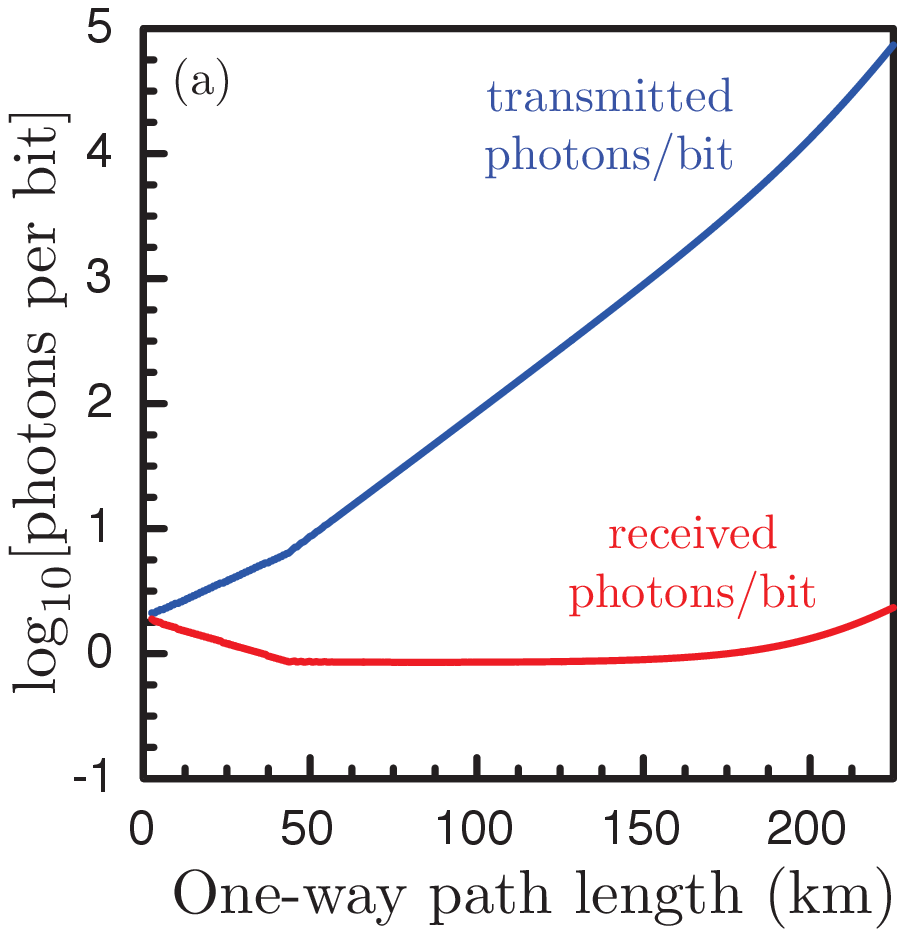}\hspace*{.05in}\includegraphics[width=1.75in]{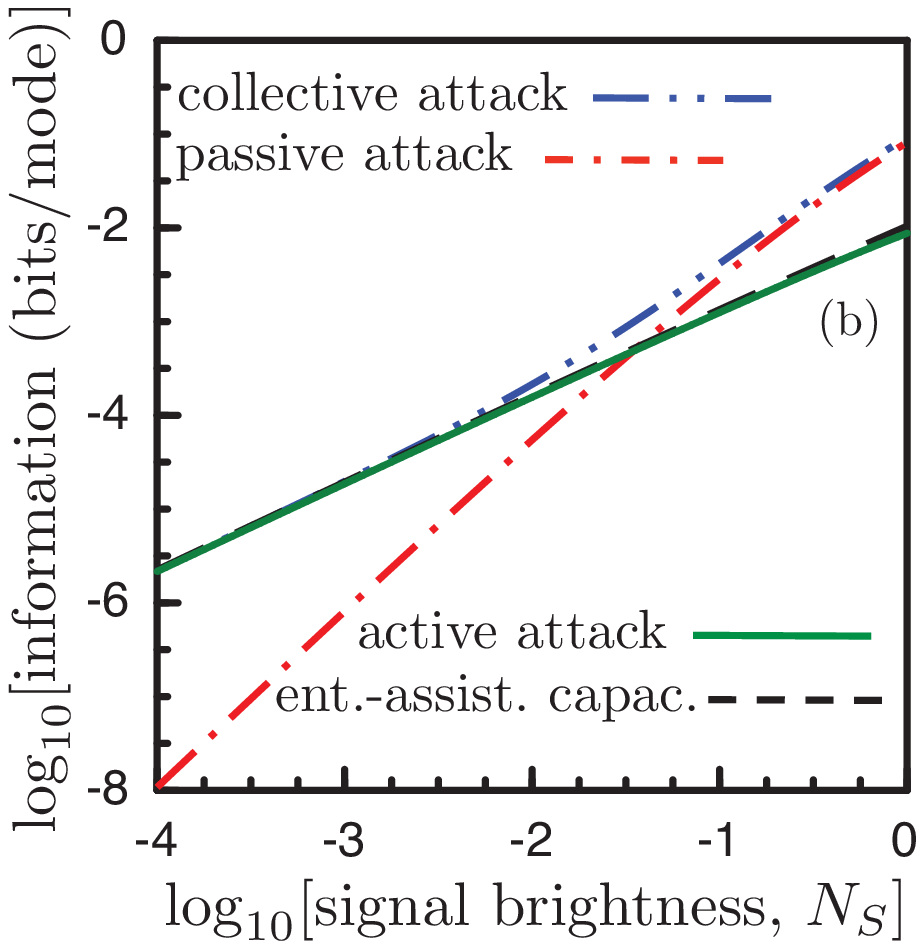}
\caption{\label{ppb}(color online). (a) Alice's transmitted photons per bit (ppb) and Bob's received ppb when Eve mounts her optimum frequency-domain collective attack with injection fraction $f_E = 0.01$.   (b) Upper bounds on Eve's optimum frequency-domain collective attack, passive attack, and active attack Holevo informations per mode---along with her entanglement-assisted capacity---as a function of Alice's signal brightness, $N_S$, for a 50\,km one-way path length assuming $f_E = 0.01$.}
\end{figure}

We see from Fig.~\ref{keyrates}(a) that 2\,Gbps QKD is possible at 50\,km one-way path length when $f_E = 0.01$, and that this secret-key rate is obtained with $N_S = 0.043$.  (Figure~\ref{keyrates}(b) shows how this rate degrades as Eve's injection fraction increases.) Thus, as suggested at the outset, security against a collective attack has been ensured by a combination of low-brightness transmission and coincidence-based channel monitoring.  That FL-QKD has such a high rate after the 10\,dB of one-way propagation loss incurred at 50\,km is then due to its use of an optical bandwidth far in excess of its modulation rate, which enables Alice to transmit many photons per bit (ppb) without affording Eve very much information.  This follows from Fig.~\ref{ppb}(a), which plots the ppb that Alice transmits to Bob and the ppb that Bob receives from Alice.  We see that FL-QKD maintains a near-unity ppb received by Bob for all path lengths less than 200\,km \cite{footnote2}.  The highest rate, long-distance, DV-QKD demonstration---Lucamarini \emph{et al}.'s BB84 system \cite{Lucamarini2013}---employs $\sim$1 \emph{transmitted} ppb.  Hence it cannot match FL-QKD's loss-independent $\sim$1 \emph{received} ppb performance.  Thus its long-distance secret-key rate is vastly inferior to FL-QKD's.  Moreover, as noted earlier, an impractically large amount of WDM would be needed for that BB84 system to match FL-QKD's single-channel Gbps secret-key rate capability over 50\,km of fiber.  

The story for Huang \emph{et al}.'s CV-QKD demonstration \cite{Huang2015} is a little different.  CV-QKD transmissions are better quantified in terms of photons per channel use rather than photons per bit, quantities that are identical for BB84 systems and for FL-QKD but typically different for CV-QKD systems.  Moreover, CV-QKD systems do not limit themselves to $\sim$1\,photon/use.  Nevertheless, even scaling it up to a 10\,Gbaud source rate, Huang \emph{et al}.'s system would still require more than 50 WDM channels to realize a 1\,Gbps secret-key rate on a 50-km-long link.   

We will close our secret-key rate assessment with some additional comments on its underlying security analysis. Consider first the optimality of Eve's using SPDC light injection in the Fig.~\ref{realization} setup.  For a given value of her injection fraction, $f_E$, Eve's use of an SPDC source in an active attack yields a Holevo information that saturates the entanglement-assisted capacity for the channel created by her injection, Bob's BPSK modulation and optical amplification, and her tap of the Bob-to-Alice channel.  Hence this confirms that no non-Gaussian active attack with the same $f_E$ can do any better. This behavior is illustrated in Fig.~\ref{ppb}(b), for a 50\,km one-way path length and $f_E = 0.01$, where we have plotted our upper bound on Eve's active-attack Holevo information per mode versus Alice's signal brightness, $N_S$, along with Eve's entanglement-assisted capacity \cite{AppC2F}.  Further insights from Fig.~\ref{ppb}(b) come from its display of Eve's passive-attack and optimum frequency-domain collective attack Holevo informations per mode \cite{AppC2}.   When $N_S \le 10^{-3}$, the active attack is almost as powerful as the optimum frequency-domain collective attack, but at $N_S \ge 0.1$ the passive attack makes the dominant contribution to the optimum frequency-domain collective attack \cite{footnote3}.  These characteristics are easily understood from the simple, individual-attack error probabilities we presented earlier.  For both passive and fixed-$f_E$ active attacks, Eve's error probability decreases with increasing $N_S$, but her passive-attack error exponent is proportional to $N_S^2$ at low brightness, whereas her fixed-$f_E$ active-attack error exponent is proportional to $N_S$.   In future work we will pursue security analysis for coherent attacks.  Because FL-QKD can be regarded as a two-way CV-QKD protocol that uses discrete modulation, coherent-attack security analyses for one-way CV-QKD \cite{Leverrier2009,Leverrier2011a,Leverrier2011b} may provide a useful starting point.  

\section{Discussion} We have argued that a QKD system's secret-key rate, in bits per second, is its preeminent figure of merit, and we have shown that single-channel FL-QKD vastly outperforms its state-of-the-art competition for long-distance OTP distribution.  To elaborate on why that is so, let us compare FL-QKD's secret-key \emph{efficiency}, in bits per channel use, with those of the highest-rate, long-distance DV-QKD and CV-QKD systems.  The secret-key efficiency of Lucamarini \emph{et al}.'s DV-QKD system at 50\,km is 1\,Mbps/1\,Gbps = $10^{-3}$\,bits/use, while the extrapolated secret-key efficiency for Huang \emph{et al}.'s CV-QKD system is 90\,kbps/50\,Mbaud = $1.8\times 10^{-3}$\,bits/use at that distance.  FL-QKD, however, is predicted to have a secret-key efficiency of 0.2\,bits/use at 50\,km, two orders of magnitude better than state-of-the-art performance.  Pirandola \emph{et al.} \cite{Pirandola2015b,footnote4} have shown that the ultimate limit for any QKD protocol's secret-key efficiency, in bits per \emph{mode}, is $-\log_2(1-\kappa_S) = 0.152$ bits per mode for a 50-km-long fiber with 0.2\,dB/km loss.  Because CV-QKD must mode-match its LO to its signal, CV-QKD's secret-key efficiencies in bits per channel use and bits per mode will coincide.  Ideal DV-QKD systems also use single-mode transmission, in which case their secret-key efficiencies in bits per channel use and bits per mode will coincide.  FL-QKD, on the other hand, employs many modes per channel use:  at 50\,km, our 10\,Gbps modulation rate and 2\,THz ASE bandwidth imply there are 200 modes per channel use, making FL-QKD's secret-key efficiency in bits per mode $0.2\,{\rm bits/use} \div 200\,{\rm modes/use} = 10^{-3}\,{\rm bits/mode}$, i.e., on par with Huang \emph{et al}.'s and Lucamarini \emph{et al}.'s.  

Before closing, two additional points need some attention.  Both are related to our use of coincidence-based channel monitoring---the first concerns what information that monitoring might reveal to Eve and the second has to do with preventing Eve from eluding that monitoring with an intercept-resend attack---and both will be part of our continuing security analysis for FL-QKD.

In their channel monitoring, Alice and Bob will record the times at which their monitors have detected photons.  Bob will transmit his detection times to Alice---over their tamper-proof classical connection---and Alice, in turn, will merge that data with her own to find the singles and coincidence rates she needs to determine the value of Eve's intrusion parameter, $f_E$.  As part of her frequency-domain collective attack, Eve can listen to Alice and Bob's classical channel, and use Bob's photon-detection information to help her decode Bob's transmission.  The security analysis we have presented thus far does not account for that possibility.  We show, however, in App.~G, that Eve's Holevo information rate increases by an inconsequential amount when she pays attention to Bob's detection-time data.  Indeed, the resolution of the secret-key rate plot in Fig.~3(a) is insufficient to show the effect.  

Although Eve's frequency-domain collective attack derives no appreciable benefit from learning the photon-detection times of Bob's channel monitor, she could take an altogether different approach to breaking FL-QKD:  an intercept-resend attack.  By detecting the photons that Alice sends to Bob, Eve could transmit her own light---with photons concentrated at those detection times---in the hope that Bob's channel-monitor data will be indistinguishable from what he would get were she not present.  Whether Eve could do so without changing Alice and Bob's $f_E$ measurement is unclear, as is whether Eve could do so while simultaneously being able to retain a suitable reference beam for decoding Bob's message, but it is important to note that intercept-resend is \emph{not} a frequency-domain collective attack, although security against it would be included were we able to prove FL-QKD's security against a general coherent attack.  Even without that coherent-attack analysis, Alice and Bob's can augment their channel monitors to at least \emph{detect} an intercept-resend attack---and hence turn it into a denial-of-service attack---by exploiting the entanglement between the signal and idler outputs of Alice's SPDC source.  Alice and Bob's coincidence-based channel monitoring only relies on the photon-paired nature of those signal and idler beams, which is why Eve could potentially duplicate that pairing.  Entanglement, on the other hand, cannot be spoofed.  So, if Alice and Bob add either dispersive-optics (frequency-domain coincidence) measurements (as in \cite{Mower2013}), or a Franson interferometer (as in \cite{Zhang2014}), to their channel monitors, it will be impossible for Eve to mount an intercept-resend attack without being detected.

In conclusion, existing single-channel QKD systems' secret-key rates at 50\,km are so low that their Gbps WDM versions have overwhelming implementation and cost issues.  With Gbps FL-QKD, however, OTP encryption of large files becomes practical over metropolitan-area networks using only a single channel.   In this regard we emphasize that FL-QKD needs \emph{no} new technology:  erbium-doped fiber amplifiers suffice for Alice's ASE source and Bob's amplifier; high-quality SPDC's are capable of the brightness that Alice requires; BPSK modulators capable of 10\,Gbps rates are readily available; Alice's receiver can use commercially available balanced mixers and need \emph{not} be shot-noise limited \cite{footnote5}; and Alice and Bob's channel monitors can employ available superconducting nanowire detectors.  

We acknowledge support from ONR grant number N00014-13-1-0774, AFOSR grant number FA9550-14-1-0052, and the DARPA Quiness Program through ARO Grant number W31P4Q-12-1-0019.  We also acknowledge the anonymous reviewer who directed us to address the information that Eve might gain from knowing the photon-detection times from Bob's channel monitor.

\begin{appendix}
\section{Alice and Bob's Terminals}
In this section we will detail the equipment that Alice and Bob use in the FL-QKD setup shown in Fig.~\ref{setup}.

\subsection{Alice's Transmitter}

Alice uses both a spontaneous parametric downconverter (SPDC) and an amplified spontaneous emission (ASE) source.  For each bit interval, the SPDC source produces $M = TW\gg 1$ signal-idler mode pairs---where $T = 1/R$ gives the bit duration in terms of Bob's modulation rate $R$, and $W$ is the SPDC's phase-matching bandwidth---with annihilation operators $\{\,(\hat{a}_{S_m}^{\rm SPDC},\hat{a}_{I_m}^{\rm SPDC}) : 1 \le m \le M\,\}$.  These SPDC mode pairs are in independent, identically-distributed, zero-mean Gaussian pure states that are characterized by the Wigner covariance matrix
\begin{equation}
{\boldsymbol \Lambda}_{SI}^{\rm SPDC} =  \frac{1}{4}\left[\begin{array}{cc}
{\bf A}_{\rm SPDC} & {\bf C}_{\rm SPDC}\\[.05in]
{\bf C}_{\rm SPDC} & {\bf A}_{\rm SPDC} \end{array}\right],
\end{equation}
where ${\bf A}_{\rm SPDC} = (2N_{\rm SPDC} +1){\bf I}_2$, 
with ${\bf I}_2$ being the $2 \times 2$ identity matrix, and
\begin{equation}
{\bf C}_{\rm SPDC} = \left[\begin{array}{cc}
C_{\rm SPDC} & 0\\[.05in]
0 & -C_{\rm SPDC} \end{array}\right],
\end{equation}
with $N_{\rm SPDC} \ll 1$ and $C_{\rm SPDC} = 2\sqrt{N_{\rm SPDC}(N_{\rm SPDC}+1)}$.  For each bit interval, the ASE source---whose $W$\,Hz bandwidth and center frequency match those of the SPDC's signal beam---produces $M$ signal-reference mode pairs, with annihilation operators $\{\,(\hat{a}^{\rm ASE}_{S_m},\hat{a}^{\rm ASE}_{R_m}) : 1\le m \le M\,\}$.  These ASE mode pairs are in independent, identically-distributed, completely-correlated thermal states that are characterized by the Wigner covariance matrix,
\begin{equation}
{\boldsymbol \Lambda}^{\rm ASE}_{SR} =  \frac{1}{4}\left[\begin{array}{cc}
{\bf A}_{\rm ASE} & {\bf C}_{\rm ASE}\\[.05in]
{\bf C}_{\rm ASE} & {\bf A}_{\rm LO} \end{array}\right],
\end{equation}
where ${\bf A}_{\rm ASE} = (2N_{\rm ASE} +1){\bf I}_2$,
${\bf C}_{\rm ASE} = 2\sqrt{N_{\rm ASE}N_{\rm LO}}\,{\bf I}_2$,
and ${\bf A}_{\rm LO} = (2N_{\rm LO} +1){\bf I}_2$, with $N_{\rm ASE} = 1 \ll N_{\rm LO}$,  

Alice sends her SPDC's idler beam to a channel monitor, and combines her SPDC and ASE source's signal beams on an asymmetric beam splitter obtaining output modes,
\begin{equation}
\hat{a}_{A_m} = \sqrt{\kappa_C}\,\hat{a}^{\rm SPDC}_{S_m} + \sqrt{1-\kappa_C}\,\hat{a}^{\rm ASE}_{S_m}.
\end{equation}
Because she wants each of these modes to have average photon number $N_A \ll 1$, and she wants their ASE-to-SPDC ratio to be $n$:1 with $n \gg 1$, Alice uses $\kappa_C = 1-nN_A/(n+1)$, and adjusts her downconverter's pump power to obtain $N_{\rm SPDC} = N_A/[n(1-N_A)+1]$.  Note that for $N_A \le 0.1$ and $n = 99$, these choices imply $\kappa_C \ge 0.9$.  

Alice now directs a fraction $\kappa_A$ of her ASE-SPDC signal light to a channel monitor and sends the remaining portion to Bob; the latter's $M$ modes are governed by annihilation operators
\begin{equation}
\hat{a}_{S_m} = \sqrt{1-\kappa_A}\,\hat{a}_{A_m} + \sqrt{\kappa_A}\,\hat{v}_{A_m},
\label{SmSA}
\end{equation}
where the noise modes $\{\hat{v}_{A_m}\}$ are in their vacuum states.  It follows that the signal modes Alice sends to Bob, her SPDC idler modes, and her ASE reference modes---i.e., the $\{\,(\hat{a}_{S_m},\hat{a}^{\rm SPDC}_{I_m},\hat{a}^{\rm ASE}_{R_m}): 1\le m \le M\,\}$---are independent, identically-distributed mode triples.  Each such mode triple is in a zero-mean Gaussian state that is completely characterized by the Wigner covariance matrix
\begin{equation}
{\boldsymbol \Lambda}_{SIR} = 
\frac{1}{4}\left[\begin{array}{ccc}
{\bf A}_S & {\bf C}'_{\rm SPDC} & {\bf C}'_{\rm ASE}\\[.05in]
{\bf C}'_{\rm SPDC} & {\bf A}_{\rm SPDC} & {\bf 0} \\[.05in]
{\bf C}'_{ASE} & {\bf 0} & {\bf A}_{\rm LO}  
\end{array}\right],
\label{covSIR}
\end{equation} 
where ${\bf A}_S = (2N_S+1){\bf I}_2 $, $N_S = (1-\kappa_A)N_A$, ${\bf C}'_{\rm SPDC} = \sqrt{(1-\kappa_A)\kappa_C}\,{\bf C}_{\rm SPDC}$, and ${\bf C}'_{\rm ASE} = \sqrt{(1-\kappa_A)(1-\kappa_C)}\,{\bf C}_{\rm ASE}$.

\subsection{Bob's Terminal}
For each bit interval, Bob receives a collection of independent, identically-distributed modes with annihilation operators $\{\,\hat{a}_{S_m}' : 1\le m \le M\}$.  He first diverts a fraction $\kappa_B$ of each mode to his channel monitor before sending the remaining light---with annihilation operators
\begin{equation}
\hat{a}''_{S_m} = \sqrt{1-\kappa_B}\,\hat{a}'_{S_m} + \sqrt{\kappa_B}\,\hat{v}_{B_m},
\end{equation}
where the noise modes $\{\hat{v}_{B_m}\}$ are in their vacuum states---to his binary phase-shift keying (BPSK) modulator.
Bob then amplifies the modulated modes with an erbium-doped fiber amplifier (EDFA) with gain $G_B$ and output ASE $N_B\ge G_B -1$.  The modes that Bob transmits to Alice therefore have photon annihilation operators
\begin{equation}
\hat{a}_{B_m} = (-1)^b\sqrt{G_B}\,\hat{a}''_{S_m} +  \sqrt{G_B-1}\,\hat{n}_{B_m}^\dagger,
\label{BobsAmp}
\end{equation}
where $b=0$ or 1 is Bob's bit value and the noise modes $\{\hat{n}_{B_m}\}$ are in independent, identically-distributed thermal states with $\langle \hat{n}_{B_m}\hat{n}^\dagger_{B_m}\rangle = N_B/(G_B-1) \ge 1$.  

\subsection{Alice's Receiver}
For a bit interval in which Bob has transmitted the value $b$, Alice receives a collection of independent, identically-distributed modes with annihilation operators $\{\,\hat{a}_{B_m}' : 1\le m \le M\,\}$.  Alice detects them using a balanced-homodyne arrangement and decides on the value of Bob's bit by comparing the outcome of that 
\begin{equation}
\hat{N}_{\rm hom} = \sum_{m=1}^M \!\left(\hat{a}^{'\dagger}_{+m}\hat{a}'_{+m} - 
\hat{a}^{'\dagger}_{-m}\hat{a}'_{-m}\right)
\end{equation}
measurement with zero. She decides that Bob sent $b=0$ if the measurement outcome exceeds zero, and she decides $b=1$ otherwise \cite{footnote6}.  In this expression,
\begin{equation}
\hat{a}'_{\pm m} = \sqrt{\eta}\left(\frac{\hat{a}'_{B_m} \pm \hat{a}'_{R_m}}{\sqrt{2}}\right) + \sqrt{1-\eta}\,\hat{v}_{\pm_m},
\end{equation}
where $\eta$ is the homodyne detector's efficiency, i.e., the product of its mode-mixing and quantum efficiencies, and the noise modes $\{\hat{v}_{\pm_m}\}$ are in their vacuum states.  

The reference modes, $\{\hat{a}_{R_m}\}$, undergo optical amplification, with gain $G_R$ and output ASE $N_R = G_R$, prior to being stored in a transmissivity-$\kappa_I$ fiber spool---whose length is chosen so that its output will be delay matched to the light Alice receives from Bob---resulting in   
\begin{equation}
\hat{a}'_{R_m} = \sqrt{\kappa_I}\left(\sqrt{G_R}\hat{a}_{R_m} + \sqrt{G_R-1}\,\hat{n}^\dagger_{R_m}\right)  + \sqrt{1-\kappa_I}\,\hat{v}_{R_m},
\end{equation}
with the amplifier-noise modes $\{\hat{n}_{R_m}\}$ being in independent, identically-distributed thermal states with $\langle \hat{n}_{R_m}\hat{n}^\dagger_{R_m}\rangle = N_R/(G_R-1)$ and the loss-noise modes $\{\hat{v}_{R_m}\}$ being in their vacuum states.  For $N_{\rm LO} \gg 1$ and $G_R = 1/\kappa_I$, this amplify-then-store procedure leaves the average photon number of the reference almost unchanged and it preserves nearly-complete correlation between the stored reference and the signal beam that Alice sent to Bob.  In particular, before storage we have that
\begin{equation}
\langle\hat{a}_{R_m}^\dagger\hat{a}_{R_m}\rangle = N_{\rm LO},
\end{equation}
and
\begin{align}
\frac{|\langle \hat{a}_{S_m}^\dagger\hat{a}_{R_m}\rangle|^2}{\langle\hat{a}_{S_m}^\dagger\hat{a}_{S_m}\rangle\langle \hat{a}_{R_m}^\dagger\hat{a}_{R_m}\rangle} &= \frac{(1-\kappa_C)N_{\rm ASE}}{\kappa_CN_{\rm SPDC} + (1-\kappa_C)N_{\rm ASE}} \nonumber \\ &= n/(n+1),
\end{align}
while after storage we find that 
\begin{align}
\langle\hat{a}^{'\dagger}_{R_m}\hat{a}'_{R_m}\rangle &= \kappa_IG_RN_{\rm LO} + \kappa_IN_R 
\nonumber \\[.05in]
&= N_{\rm LO} + 1 \approx N_{\rm LO},
\end{align}
and
\begin{eqnarray}
\lefteqn{\frac{|\langle \hat{a}_{S_m}^\dagger\hat{a}'_{R_m}\rangle|^2}{\langle\hat{a}_{S_m}^\dagger\hat{a}_{S_m}\rangle\langle \hat{a}_{R_m}^{'\dagger}\hat{a}'_{R_m}\rangle} } \nonumber \\[.05in] 
&=& \frac{(1-\kappa_C)N_{\rm ASE}N_{\rm LO}}{(\kappa_CN_{\rm SPDC} + (1-\kappa_C)N_{\rm ASE})(N_{\rm LO}+1)} \nonumber \\[.05in] &\approx& n/(n+1),
\end{eqnarray}
when $N_{\rm LO} \gg 1$ \cite{footnote7}.  For $n=99$, as assumed in the paper's secret key-rate calculations, we see that Alice's reference suffers almost no degradation.

\section{Eve's Frequency-Domain Collective Attack}
\begin{figure}[htb]
\includegraphics[width=3.25in]{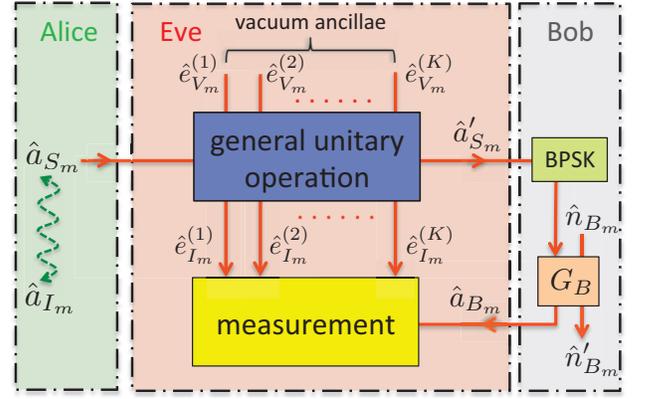}
\caption{\label{attack}Schematic of Eve's $K$-mode collective attack used to upper bound her Holevo information rate.  BPSK: binary phase-shift keying.  $G_B$ amplifier gain.  The dashed wavy line represents an entanglement that purifies the state of the $\hat{a}_{S_m}$ mode.}
\label{scheme_general}
\end{figure}

Figure~\ref{attack} shows the structure of Eve's general frequency-domain collective attack that we will use to place an upper bound on her Holevo information rate.  Eve has replaced the low-loss (0.2\,dB/km) fibers that Alice and Bob presume are connecting their terminals with lossless fibers.  For each of Alice's $M$ transmitted modes, $\{\,\hat{a}_{S_m} : 1\le m \le M\,\}$, in a bit interval, Eve then performs the same general unitary operation on $K$ ancilla modes, $\{\,\hat{e}^{(k)}_{V_m} : 1 \le k \le K\,\}$, and Alice's $\hat{a}_{S_m}$, resulting in Bob's receiving the $\hat{a}_{S_m}'$ mode.
Here, without loss of generality, we will assume that the $\{\hat{e}^{(k)}_{V_m}\}$ are in their vacuum states.
 
For each bit interval, Eve retains the $KM$ ancilla output modes, $\{\,\hat{e}_{I_m}^{(k)} : 1\le k \le K, 1\le m \le M\,\}$, from her unitary operation and the light she taps from the Bob-to-Alice channel in a quantum memory.  At the end of the key distribution session she then makes a collective measurement in her attempt to capture all of Bob's bit values.  Because we will derive only an upper bound on Eve's Holevo information rate from this procedure, Fig.~\ref{attack} shows Eve as taking \emph{all} the light Bob sends to Alice.  Other concessions to Eve that will be used in obtaining our upper bound are:  (1)  Bob will not divert any light to his channel monitor, i.e., $\kappa_B = 0$; and (2) Bob's amplifier will have quantum-limited ASE, viz., $N_B = G_B-1$.  All of these conditions increase Eve's Holevo information rate.  That said, in practice Eve will \emph{not} collect all the light that Bob sends to Alice, Bob \emph{will} do channel monitoring ($\kappa_B >0$), and Bob's amplifier may \emph{not} be quantum limited ($N_B > G_B-1$).  Furthermore, in order to minimize Alice's ability to detect Eve's presence by simple photon-flux and spectrum monitoring, Eve will not inject any of her own light into Alice's receiver and she will arrange that the Bob-to-Alice channel still has transmissivity $\kappa_S = 10^{-0.02L}$ that Alice and Bob expect.  

\section{Upper Bound on Eve's Holevo Information Rate}
Let $\hat{\bf e}_I$ denote $\{\,\hat{e}_{I_m}^{(k)} : 1\le k \le K, 1\le m \le M\,\}$ and $\abold_B$ denote $\{\,\hat{a}_{B_m} : 1\le m \le M\,\}$. Eve's Holevo information rate for her general frequency-domain collective attack is bounded above by 
\begin{equation}
\chi_{EB} =  R\left[S(\hat{\rho}_{\hat{\bf e}_I,\abold_B}) - \sum_{b=0}^1S(\hat{\rho}_{\hat{\bf e}_I,\abold_B}^{(b)})/2\right],
\label{upperbound1}
\end{equation}
where $S(\cdot)$ denotes von Neumann entropy, $\hat{\rho}_{\hat{\bf e}_I,\abold_B}^{(b)}$ is the conditional joint density operator for the 
$\hat{\bf e}_I$ and $\abold_B$  modes given Bob's bit value, $\hat{\rho}_{\hat{\bf e}_I,\abold_B} = \sum_{b=0}^1\hat{\rho}_{\hat{\bf e}_I,\abold_B}^{(b)}/2$ is their unconditional joint density operator, and the bound is due to our assuming that Eve captures all the light Bob sends to Alice.  

Before going into details, we place two constraints on Eve's attack. First, we assume that Eve precludes her presence being detected from simple photon-flux and spectrum monitoring at Bob's terminal by requiring her attack to satisfy
\begin{equation}
\langle \hat{a}_{S_m}'^\dagger\hat{a}_{S_m}'\rangle = \kappa_SN_S,
\label{mean_photon_constraint}
\end{equation}
where $\kappa_S  = 10^{-0.02L}$ is the transmissivity of the $L$-km-long connection Alice and Bob believe they have and $N_S$ is the brightness of the light Alice sends to Bob. Second, Alice and Bob's channel monitors allow them to measure Eve's intrusion parameter, $f_E$, that,  as shown in App.~E, measures Eve's degradation of the phase-sensitive cross covariance between Alice's $\hat{a}_{I_m}^{\rm SPDC}$ mode and Bob's $\hat{a}'_{S_m}$ mode.  Because $\hat{a}_{I_m}$ is a purification of $\hat{a}_{S_m}$, it follows that there is a $0<\kappa'<1$ such that $\sqrt{\kappa'}\,\hat{a}_{I_m}$ has the same phase-sensitive cross covariance with $\hat{a}_{S_m}$ as does $\hat{a}_{I_m}^{\rm SPDC}$, so we have that
\begin{equation}
{|\braket{\hat{a}_{S_m}'\hat{a}_{I_m}}|^2}=(1-f_E)\kappa_S|\braket{\hat{a}_{S_m}\hat{a}_{I_m}}|^2 \label{fE_cov2}.
\end{equation}
Equations~(\ref{mean_photon_constraint}) and (\ref{fE_cov2}) both constrain what Eve's general frequency-domain collective attack does to the Wigner covariance matrix of the $(\hat{a}_{S_m}', \hat{a}_{I_m})$ mode pair.

To proceed further, we first introduce $\abold_I = \{\,\hat{a}_{I_m} : 1 \le m \le M\}$ that purifies $\abold_S=\{\,\hat{a}_{S_m} : 1 \le m \le M\}$, i.e., the mode pairs $\{\,(\hat{a}_{S_m},\hat{a}_{I_m}) : 1\le m \le M\,\}$ are in independent, identically-distributed, zero-mean Gaussian pure states that are characterized by the Wigner covariance matrix
\begin{equation}
{\boldsymbol \Lambda}_{SI} =  \frac{1}{4}\left[\begin{array}{cc}
{\bf A}_S & {\bf C}_S \\
{\bf C}_S & {\bf A}_S \end{array}\right],
\end{equation}
where 
\begin{equation}
{\bf C}_S = \left[\begin{array}{cc} 2\sqrt{N_S(N_S + 1)} & 0 \\[.05in] 0 & -2\sqrt{N_S(N_S+1)} \end{array}\right].
\end{equation} 
After Eve's unitary operation, however, the $\{\hat{a}_{S_m}^\prime\}$ modes will, in general, be in non-Gaussian states. Next, we introduce the complement to the Eq.~(\ref{BobsAmp}) input-output relation for Bob's amplifier, i.e., 
\begin{equation}
\hat{n}_{B_m}' = \sqrt{G_B}\,\hat{n}_{B_m} + (-1)^b\sqrt{G_B-1}\,\hat{a}_{S_m}'',
\end{equation}
with $\hat{a}_{S_m}'' = \hat{a}_{S_m}'$ because our upper bound will be found using $\kappa_B = 0$, and $\hat{n}_{B_m}$ in its vacuum state because that bound will presume Bob's amplifier is quantum limited.  With these assumptions, we have that the $\{\,\abold_I,\abold_S,\hat{\bf e}_V,\hat{\bf n}_B\,\}$ modes---where $\hat{\bf e}_V = \{\,\hat{e}_{V_m}^{(k)} : 1\le k \le K,1 \le m \le M\}$, and $\hat{\bf n}_B = \{\,\hat{n}_{B_m} : 1 \le m \le M\}$---are in a zero-mean Gaussian pure state.  It then follows that the $\{\,\abold_I,\abold_B,\hat{\bf e}_I,\hat{\bf n}_B'\,\}$ modes---where $\hat{\bf n}_B' = \{\,\hat{n}_{B_m}' : 1 \le m \le M\}$---are in a (not necessarily zero-mean Gaussian) pure state given Bob's bit value, because Eve and Bob's operations are unitary.  An immediate consequence of this purity is
\begin{equation}
S(\hat{\rho}_{\hat{\bf e}_I,\abold_B}^{(b)}) =  S(\hat{\rho}_{\abold_I,\hat{\bf n}_B'}^{(b)}).
\end{equation}
Moreover, the unitarity of the phase modulation that Bob performs, given his bit value, implies that these conditional entropies are independent of $b$.   So, 
because the mode pairs $\{\,\hat{a}_{I_m},\hat{n}_{B_m}' : 1 \le m \le M\,\}$ are in independent, identically-distributed states given Bob's bit value, we have that
\begin{equation}
\sum_{b=0}^1S(\hat{\rho}_{\abold_I,\hat{\bf n}_B'}^{(b)})/2 = MS(\hat{\rho}_{\hat{a}_{I_m},\hat{n}_{B_m}'}^{(0)}).
\end{equation}

Having obtained a simplified expression for the second entropy term on the right in (\ref{upperbound1}), we use the subadditivity of von Neumann entropy to get
\begin{equation}
\chi_{EB}\le R \left[S(\hat{\rho}_{\hat{\bf e}_I}) + S(\hat{\rho}_{\abold_B}) -M S(\hat{\rho}_{\hat{a}_{I_m},\hat{n}_{B_m}'}^{(0)})\right], 
\end{equation}
with equality when $\hat{\rho}_{\hat{\bf e}_I\abold_B}=\hat{\rho}_{\hat{\bf e}_I}\otimes \hat{\rho}_{\abold_B}$.
The $\{\hat{\bf e}_I\}$ modes are independent of Bob's bit value.  Grouping them by mode index $m$, i.e., writing $\{\hat{\bf e}_I\} = \{\,\hat{\bf e}_{I_m}: 1\le m \le M\,\}$ where $\hat{\bf e}_{I_m} = \{\,\hat{e}_{I_m}^{(k)} : 1\le k \le K\,\}$, we have that the $\{\hat{\bf e}_{I_m}\}$ modes are independent and identically distributed, so
\begin{equation}
S(\hat{\rho}_{\hat{\bf e}_I}) = M S(\hat{\rho}_{\hat{\bf e}_{I_m}}).
\end{equation}
Moreover, because Eve's operation is unitary, the $\{\hat{\bf e}_{I_m},\hat{a}_{I_m},\hat{a}_{S_m}'\}$ modes are in a pure state, so we have
\begin{equation}
S(\hat{\rho}_{\hat{\bf e}_{I_m}}) = S(\hat{\rho}_{\hat{a}_{I_m},\hat{a}_{S_m}'}).
\end{equation}
Finally, since we are considering Eve's frequency-domain collective attack, the $\{\hat{a}_{B_m}\}$ modes are independent and identically distributed, thus subadditivity gives us 
\begin{equation}
S(\hat{\rho}_{\abold_B}) \le MS(\hat{\rho}_{\hat{a}_{B_m}}).
\label{rhoABbound}
\end{equation}

Putting the preceding results together gives us an upper bound on Eve's Holevo information rate:
\begin{widetext} 
\begin{equation}
\chi_{EB} \le 
 R\min\!\left\{M\!\left[S(\hat{\rho}_{\hat{a}_{B_m}})-[S(\hat{\rho}_{\hat{a}_{I_m},\hat{n}_{B_m}'}^{(0)})-S(\hat{\rho}_{\hat{a}_{I_m},\hat{a}_{S_m}'})]\right],1\right\},
\label{upperbound2}
\end{equation}
\end{widetext}
where we have used the fact that Eve's maximum Holevo information per bit interval is one.
Our next step is to place a lower bound on $S(\hat{\rho}_{\hat{a}_{I_m},\hat{n}_{B_m}'}^{(0)})-S(\hat{\rho}_{\hat{a}_{I_m},\hat{a}_{S_m}'})$ by recognizing that term as the entropy output of a tensor-product quantum channel.\\
 
\noindent{\bf Definition:  Entropy output}\\
\emph{Let $\phi(\cdot)$ be a quantum channel that maps states in $\mathcal{H}_1$ to states in $\mathcal{H}_2$.  The entropy-output function $E_\phi(\cdot)$ of that channel  quantifies the difference between the von Neumann entropies of its output and input states, i.e., for input-state $\hat{\rho}$ we have that 
\begin{equation}
E_\phi (\hat{\rho})=S[\phi(\hat{\rho})]-S(\hat{\rho}).
\label{Entropy_output}
\end{equation}}
Using this definition (\ref{upperbound2}) can be rewritten as
\begin{equation}
\chi_{EB} \le R\min\!\left\{M[S(\hat{\rho}_{\hat{a}_{B_m}})-E_\phi(\hat{\rho}_{\hat{a}_{I_m},\hat{a}_{S_m}'})],1\right\}.
\label{upperbound_output}
\end{equation}

Next, we prove that entropy output is superadditive for the quantum channel $\phi(\cdot) = \phi_S(\cdot)\otimes I_I(\cdot)$ that maps the $\{\hat{a}'_{S_m},\hat{a}_{I_m}\}$ modes into the $\{\hat{n}'_{B_m},\hat{a}_{I_m}\}$ modes, where $I_I(\cdot)$ is the identity channel.\\

\noindent{\bf Theorem:  Superadditivity of entropy output}\\
\emph{Let $A_{12}$ and $B_{12}$ be bipartite quantum systems on $\mathcal{H}_A^{\otimes 2}$ and $\mathcal{H}_B^{\otimes 2}$ with components $\{A_1,A_2\}$ and $\{B_1,B_2\}$, respectively.  For an arbitrary input state $\hat{\rho}_{A_{12},B_{12}}$ in $\mathcal{H}_A^{\otimes 2}\otimes\mathcal{H}_B^{\otimes 2}$, and an arbitrary quantum channel $\phi(\cdot)$ that maps states in $\mathcal{H}_A\otimes \mathcal{H}_B$ into states in $\mathcal{H}_A'\otimes \mathcal{H}_B'$ we have that
\begin{equation}
E_{\phi\otimes \phi}(\hat{\rho}_{A_{12},B_{12}})\ge  E_\phi (\hat{\rho}_{A_{1},B_{1}})+E_\phi (\hat{\rho}_{A_{2},B_{2}}),
\label{superadditivity}
\end{equation}
with equality when $\hat{\rho}_{A_{12},B_{12}}=\hat{\rho}_{A_{1}B_{1}} \otimes\hat{\rho}_{A_{2}B_{2}}$, i.e., entropy output is superadditive.}\\
From entropy output's definition, Ineq.~(\ref{superadditivity}) is equivalent to
\begin{align}
&S[\phi\otimes \phi\,(\hat{\rho}_{A_{12},B_{12}})-S(\hat{\rho}_{A_{12},B_{12}})\ge S[\phi(\hat{\rho}_{A_{1},B_{1}})]&
\nonumber\\
&-S(\hat{\rho}_{A_{1},B_{1}})
+S[\phi(\hat{\rho}_{A_{2},B_{2}})]-S(\hat{\rho}_{A_{2},B_{2}}).&
\label{rewritesuperadditivity}
\end{align}
This inequality can be rewritten as
\begin{equation}
I(A_{1}B_{1}\!:\!A_{2}B_{2}) \ge I[\phi(A_{1}B_{1})\!:\!\phi(A_{2}B_{2})],
\label{mutualinfo}
\end{equation} 
where $I(A\!:\!B)=S(\hat{\rho}_{A})+S(\hat{\rho}_{B})-S(\hat{\rho}_{A,B})$ is the quantum mutual information. The validity of Ineq.~(\ref{mutualinfo}) follows from the quantum data-processing inequality \cite{quantum_data_processing}, because $\phi(\cdot)$ acts independently on $(A_1,B_1)$ and $(A_2,B_2)$.  

The subadditivity of von Neumann entropy and the superadditivity of entropy output imply that $S(\hat{\rho}_{\hat{a}_{B_m}})-E_\phi(\hat{\rho}_{\hat{a}_{I_m},\hat{a}_{S_m}'})$ is subadditive.  Moreover, von Neumann entropy is continuous.  So, if we can show that entropy output for Gaussian channels is invariant under passive symplectic operations then we could apply Gaussian extremality \cite{Wolf_2006} and obtain 
\begin{eqnarray}
\lefteqn{S(\hat{\rho}_{\hat{a}_{B_m}})-E_\phi(\hat{\rho}_{\hat{a}_{I_m},\hat{a}_{S_m}'}) \le } \nonumber \\[.05in]
&& S_G({\boldsymbol \Lambda}_{B})-[S_G({\boldsymbol \Lambda}_{IB'}^{(0)})-S_G({\boldsymbol \Lambda}_{IS'})],
\label{extremality}
\end{eqnarray}
where $S_G({\boldsymbol \Lambda})$ denotes the von Neumann entropy of a Gaussian state with Wigner covariance matrix ${\boldsymbol \Lambda}$, and ${\boldsymbol \Lambda}_B$, ${\boldsymbol \Lambda}_{IB'}^{(0)}$, and ${\boldsymbol \Lambda}_{IS'}$ are the Wigner covariance matrices of $\hat{\rho}_{\hat{a}_{B_m}}$, $\hat{\rho}_{\hat{a}_{I_m},\hat{n}_{B_m}'}^{(0)}$, and $\hat{\rho}_{\hat{a}_{I_m},\hat{a}_{S_m}'}$, respectively.  It would then follow that, Eve's Holevo information rate for her general frequency-domain collective attack satisfies
\begin{equation}
\chi_{EB}\le R \min\!\left\{M[S_G({\boldsymbol \Lambda}_{B})+ S_G({\boldsymbol \Lambda}_{IS'})-S_G({\boldsymbol \Lambda}_{IB'}^{(0)})], 1 \right\},
\label{upperbound_cov}
\end{equation}
which means that we only need to maximize this rate when Eve makes a collective frequency-domain Gaussian attack.  Note that ${\boldsymbol \Lambda}_{B}$ and ${\boldsymbol \Lambda}_{IB'}^{(0)}$ are obtained from ${\boldsymbol \Lambda}_{IS'}$ by applying Bob's modulator and amplifier transformations, and that  Eqs.~(\ref{mean_photon_constraint}) and (\ref{fE_cov2}) place constraints on ${\boldsymbol \Lambda}_{IS'}$ when Eve mounts her frequency-domain collective attack.  The rest of this section is devoted to:  (1) proving that entropy output for Gaussian channels is invariant under passive symplectic transformations; and (2) placing an explicit upper bound on Eve's Holevo information rate for her optimum frequency-domain collective Gaussian attack under the preceding covariance constraints.

To show that entropy output for Gaussian channels is invariant under passive symplectic transformations, we rely on the fact that Gaussian channels and symplectic transformations are both linear Bogoliubov mode transformations.  Also, because the $\{\hat{a}_{I_m}\}$ modes are in Gaussian states, we only need to consider symplectic transformations of the $\{\hat{a}'_{S_m}\}$ modes.  Consider a Gaussian channel $\phi_G(\cdot)$ whose input modes are $\hat{a}_1$ and $\hat{a}_2$ and whose output modes satisfy 
\begin{align}
\hat{b}_1&=c_1 \hat{a}_1+c_2 \hat{a}_1^\dagger+c_3\hat{n}_1+c_4\hat{n}_1^\dagger,\\[.05in]
\hat{b}_2&=c_1 \hat{a}_2+c_2 \hat{a}_2^\dagger+c_3\hat{n}_2+c_4\hat{n}_2^\dagger,
\end{align}
where the $\{c_k\}$ are complex-valued coefficients associated with $\phi_G(\cdot)$ and the $\{\hat{n}_k\}$ are vacuum-state ancilla modes. Now suppose that the input modes are applied to the input ports of a 50--50 beam splitter whose outputs, 
\begin{equation}
\hat{a}_\pm=(\hat{a}_1\pm \hat{a}_2)/\sqrt{2},
\end{equation} 
become the inputs to $\phi_G(\cdot)$.  Now the output modes will be 
\begin{align}
\hat{b}_+&=c_1 \hat{a}_++c_2 \hat{a}_+^\dagger+c_3\hat{n}_1+c_4\hat{n}_1^\dagger,\\[.05in]
\hat{b}_-&=c_1 \hat{a}_-+c_2 \hat{a}_-^\dagger+c_3\hat{n}_2+c_4\hat{n}_2^\dagger.
\end{align} 
Because unitary operations do not change von Neumann entropy, we can apply another 50--50 beam splitter to these output modes and obtain
\begin{align} 
\hat{b}_1^\prime&=(\hat{b}_++\hat{b}_-)/\sqrt{2},\\[.05in]
\hat{b}_2^\prime&=(\hat{b}_+-\hat{b}_-)/\sqrt{2} 
\end{align}
whose von Neumann entropy will be the same as that of the $\{\hat{b}_+,\hat{b}_-\}$ modes. With some algebra, we can verify that
\begin{eqnarray}
\hat{b}_1^\prime=c_1 \hat{a}_1+c_2 \hat{a}_1^\dagger+c_3\hat{n}_++c_4\hat{n}_+^\dagger,\\
\hat{b}_2^\prime=c_1 \hat{a}_2+c_2 \hat{a}_2^\dagger+c_3\hat{n}_-+c_4\hat{n}_-^\dagger,
\end{eqnarray}
where the $\hat{n}_\pm=(\hat{n}_1\pm \hat{n}_2)/\sqrt{2}$ are in their vacuum states. Hence the $\{\hat{b}_1^\prime,\hat{b}_2^\prime\}$ modes have the same von Neumann entropy as $\{\hat{b}_1,\hat{b}_2\}$ modes.  A similar analysis will demonstrate entropy invariance for waveplate transformations, completing the proof that the entropy output for Gaussian channels is invariant under passive symplectic transformations.

Having shown the last condition we needed for Gaussian extremality to hold, we turn our attention to Eve's collective frequency-domain Gaussian attack.  In such an attack, Eve's unitary operation in Fig.~\ref{scheme_general} is a $K+1$-mode Bogoliubov transformation~\cite{Weedbrook2012}, resulting in 
\begin{equation}
\hat{a}_{S_m}' = u_0\hat{a}_{S_m} + v_0^*\hat{a}_{S_m}^\dagger + \sum_{k=1}^K (u_k\hat{e}_{V_m}^{(k)} + v_k^*\hat{e}_{V_m}^{(k)\dagger} )+ \alpha.
\label{Bogoliubov}
\end{equation}
A direct consequence of Gaussian extremality is that the optimum displacement is $\alpha=0$, because only when $\alpha=0$ will the unconditional state $\hat{\rho}_{\hat{a}_{B_m}}$ be Gaussian.  So, setting $\alpha =0$, we need to maximize the right-hand side of 
Ineq.~(\ref{upperbound_cov}) over the parameters $\{\,u_k, v_k : 0\le k \le K\}$ subject to the following constraints. 

First, so that Eq.~(\ref{Bogoliubov}) yields a proper free-field commutator bracket for $\hat{a}_{S_m}'$, we require that the coefficients $\{\, u_k,v_k : 0 \le k \le K\,\}$ satisfy
\begin{equation}
\sum_{k=0}^K(|u_k|^2 - |v_k|^2) = 1.
\label{commutatorpreservation}
\end{equation}
Second, the security-monitoring constraint in Eq.~(\ref{mean_photon_constraint}) implies that Eve's attack parameters $\{\, u_k,v_k : 0 \le k \le K\,\}$ must obey
\begin{equation}
(|u_0|^2 + |v_0|^2)N_S + \sum_{k=0}^K|v_k|^2 = \kappa_SN_S.
\label{fluxpreservation}
\end{equation}
Because the first term on the left is Alice's light injection into Bob while the second terms is due to Eve, the constraint in Eq.~(\ref{fE_cov2}) can be rewritten as 
\begin{equation}
f_E = \frac{\sum_{k=0}^K|v_k|^2 }{\kappa_SN_S},
\label{fEconstraint}
\end{equation}
which shows that under Eve's collective frequency-domain Gaussian attack the intrusion parameter $f_E$ equals the fraction of light entering Bob's terminal that is due to Eve.  In App.~E we will show that Alice and Bob's photon-coincidence channel monitoring can measure $f_E$.  Hence Eve will constrain her attack parameters to yield an $f_E$ value that Alice and Bob will tolerate in the FL-QKD protocol.  (Eve's using an $f_E$ value that  exceeds what Alice and Bob will tolerate would constitute a denial-of-service attack.) 

\subsection{Evaluating Eve's Holevo Information Rate Upper Bound}
We can evaluate the bound in (\ref{upperbound_cov}) by symplectic diagonalization of the Wigner covariance matrices of $\{\,\hat{a}_{I_m},\hat{n}_{B_m}'\}$, $\{\hat{a}_{I_m},\hat{a}_{S_m}'\}$, and $\hat{a}_{B_m}$ conditioned on the value of Bob's bit.  From ~App.~B we can easily show that
\begin{equation}
{\boldsymbol \Lambda}_{IS'} = \frac{1}{4}\!\left[\begin{array}{ccc} {\bf A}_S & & {\bf C}_{IS'} \\[.05in]
{\bf C}_{IS'} & & {\bf B}_{IS'}\end{array}\right],
\end{equation}
where 
\begin{equation}
{\bf B}_{IS'} =  2\!\left[\begin{array}{cc} B + {\rm Re}(w)&  
{\rm Im}(w)\\[.05in]
{\rm Im}(w)&  B - {\rm Re}(w) 
\end{array}\right],
\end{equation}
and
\begin{equation}
{\bf C}_{IS'} = 2\sqrt{N_S(N_S+1)}\left[\begin{array}{ccc} {\rm Re}(u_0+v_0) & & {\rm Im}(u_0 - v_0) \\[.05in]
{\rm Im}(u_0 + v_0) & & -{\rm Re}(u_0-v_0)\end{array}\right], 
\end{equation}
with $B = 1/2 + \kappa_SN_S$, $w={\bf v}^\dagger{\bf u} + (2N_S+1)v_0^*u_0$, ${\bf v}^\dagger \equiv \left[\begin{array}{cccc} v_1^* & v_2^* & \cdots & v_K^*\end{array}\right]$ and ${\bf u} = \left[\begin{array}{cccc} u_1 & u_2 & \cdots & u_K\end{array}\right]^T$ and $^T$ denoting transpose. We also find that 
\begin{equation}
{\boldsymbol \Lambda}^{(b)}_{IB'} = \frac{1}{4}\!\left[\begin{array}{ccc} {\bf A}_S & & {\bf C}^{(b)}_{IB'} \\[.05in]
{\bf C}^{(b)}_{IB'} & & {\bf B}_{IB'}\end{array}\right],
\end{equation}
where
\begin{eqnarray}
{\bf B}_{IB'} &=&  \left[\begin{array}{ccc}B' + {\rm Re}(x)  & & -{\rm Im}(x) \\[.05in]
-{\rm Im}(x)& & B' -{\rm Re}(x) \end{array}\right],\\[.05in]
{\bf C}_{IB'}^{(b)} &=& (-1)^b2\sqrt{(G_B-1)N_S(N_S+1)} \nonumber \\[.05in]
&\times& \left[\begin{array}{ccc} {\rm Re}(u_0+v_0) & &  -{\rm Im}(u_0 - v_0) \\[.05in]
{\rm Im}(u_0 + v_0) & & {\rm Re}(u_0-v_0)\end{array}\right], 
\end{eqnarray} 
with $B' = 1+2(G_B-1)(\kappa_SN_S + 1)$ and $x = 2(G_B-1)w$.  The last Wigner covariance that we need is 
\begin{equation}
{\boldsymbol \Lambda}_B^{(b)} = \frac{1}{4}\!\left[\begin{array}{ccc} B''+ 2G_B{\rm Re}(w)& & 2G_B{\rm Im}(w)\\[.05in]
2G_B{\rm Im}(w) & & B'' - 2G_B{\rm Re}(w)\end{array}\right],
\end{equation}
where $B'' = -1 + 2G_B(\kappa_SN_S + 1 )$. Because this covariance matrix is independent of $b$, we have ${\boldsymbol \Lambda}_B={\boldsymbol \Lambda}_B^{(b)}$ and the unconditional state of $\hat{a}_{B_m}$ is Gaussian.

After evaluating all the symplectic eigenvalues of the preceding Wigner covariances, we have that
\begin{eqnarray}
\chi_{EB} &\le& R\min\!\left\{M\left[g\!\left(\frac{4\xi_{IS'+}-1}{2}\right) + g\!\left(\frac{4\xi_{IS'-}-1}{2}\right) \right.\right. \nonumber \\[.05in]
&+& g\!\left(\frac{4\xi_B-1}{2}\right) - g\!\left(\frac{4\xi_{IB'+}-1}{2}\right) \nonumber \\[.05in]
&-& \left.\left. g\!\left(\frac{4\xi_{IB'-}-1}{2}\right)\right],1\right\},
\label{upperbound3}
\end{eqnarray}
where $g(x) = (x+1)\log_2(x+1)-x\log_2(x)$ is the von Neumann entropy of a thermal state with average photon number $x$.  Here $\xi_{IS'+} \ge \xi_{IS'-}$ and $\xi_{IB'+}\ge\xi_{IB'-}$ are, respectively, the symplectic eigenvalues of ${\boldsymbol \Lambda}_{IS'}$ and ${\boldsymbol \Lambda}^{(b)}_{IB'}$, and $\xi_B$ is the symplectic eigenvalue of ${\boldsymbol \Lambda}_B$.

Because FL-QKD operates with $N_B\gg 1$, we shall replace (\ref{upperbound3}) with its leading-order expansion in that regime, namely
\begin{eqnarray}
\chi_{EB} &\le& R\min\!\left\{M\left[g(2\xi_{IS'+}-1/2) + g(2\xi_{IS'-}-1/2)\right.\right.  \nonumber \\[.05in]
&-& \left.\left. g(2\tilde{\xi}_{IB'-}-1/2) + O(N_B^{-1/2})\right], 1\right\},
\label{upperbound4}
\end{eqnarray}
where $\xi_{IS'\pm}$ is independent of $N_B$ and $\tilde{\xi}_{IB'-}$ is the $N_B \gg 1$ leading-order, $O(1)$, approximation to $\xi_{IB'-}$.  Our next task is to maximize the right-hand side of (\ref{upperbound4}) over all possible values of Eve's attack parameters, $\{\,u_k,v_k : 0 \le k \le K\,\}$, subject to the commutator-preservation constraint (\ref{commutatorpreservation}), the photon-flux constraint (\ref{fluxpreservation}), and the injection-fraction constraint (\ref{fEconstraint}).  The first of these constraints is an absolute requirement on frequency-domain collective Gaussian attacks, the second is set by Eve's desire to elude Bob's detecting her by simple photon-flux and spectrum monitoring, and the third is a consequence of Alice and Bob's photon-coincidence monitoring.

The preceding attack-parameter optimization can be accomplished more readily by satisfying (\ref{commutatorpreservation}), (\ref{fluxpreservation}), and (\ref{fEconstraint}) by means of
\begin{align}
|v_0| &= \sqrt{(1-f_E)\kappa_S}\,\cos(\gamma_v),\nonumber\\
& \mbox{with $\gamma_v\in[0,\pi/2]$ and $\cos^2(\gamma_v) \le f_EN_S/(1-f_E)$} \label{v0}\\[.05in]
|u_0| &= \sqrt{(1-f_E)\kappa_S}\,\sin(\gamma_v) \label{u0}\\[.05in]
{\bf v}^\dagger{\bf v} &= [f_E\kappa_SN_S -(1-f_E)\kappa_S\cos^2(\gamma_v)]\label{vnorm}\\[.05in]
{\bf u}^\dagger{\bf u} &= f_E\kappa_SN_S + 1-(1-f_E)\kappa_S\nonumber \\ &+ (1-f_E)\kappa_S\cos^2(\gamma_v),\label{unorm}\\[.05in]
|{\bf v}^\dagger{\bf u}| &= \sqrt{({\bf v}^\dagger{\bf v})({\bf u}^\dagger{\bf u})}\,\cos(\delta),\mbox{ with $\delta\in[0,\pi/2]$.}\label{vu}
\end{align}
Next, we further simplify (\ref{upperbound4}) by restricting it to FL-QKD's desired long-distance operating regime, wherein $\kappa_S \ll 1$.  Here we find that
\begin{align}
&\chi_{EB} \le R\min\!\left(M\left\{\kappa_S[f_EN_S - (1-f_E)\cos^2(\gamma_v)]\sin^2(\delta)\right. \right.&
\nonumber \\
&\times \{1/\ln(2) - \log_2[\sin^2(\delta)\kappa_S[f_EN_S -(1-f_E)\cos^2(\gamma_v)]]\} \nonumber& \\
&+ 
(1-f_E)\kappa_S\log_2(1+1/N_S)[(2N_S+1)\cos^2(\gamma_v) + N_S^2]&
\nonumber\\
&+ \left. \left.O(\kappa_S^{3/2}) + O(N_B^{-1/2})\right\},1\right).&
\label{upperbound5}
\end{align}
Neglecting the $O(\cdot)$ terms, we find that the derivative of the right-hand side of (\ref{upperbound5}) with respect to $\sin^2(\delta)$ will be positive if $\ln[2f_E\kappa_SN_S] <0$, a condition that will always be satisfied when $\kappa_SN_S \ll 1$.  Thus we conclude that $\delta = \pi/2$ is Eve's best choice. Next, using $\delta= \pi/2$ in (\ref{upperbound5}), neglecting the $O(\cdot)$ terms, and differentiating  (\ref{upperbound5})'s right-hand side with respect to $\cos^2(\gamma_v)$, we find that it will be negative if 
\begin{align}
\ln(2f_E\kappa_S) &< -\max_{N_S \le 1}[\ln(N_S) + (1+2N_S)\ln(1+1/N_S)] \nonumber \\ 
& \approx -2,
\label{condition}
\end{align}
where the $N_S$ constraint is due to FL-QKD's operating at low brightness. Alice and Bob's constraining Eve to $f_E \ll 1$ combined with $\kappa_S \ll 1$ ensures that (\ref{condition}) is obeyed, making $\gamma_v = \pi/2$ optimum.  

At this point, using $\delta = \gamma_v = \pi/2$ in Eqs.~(\ref{u0})--(\ref{vu}), we have that Eve's optimum frequency-domain collective Gaussian attack is to use the Fig.~\ref{attack} setup with 
\begin{align}
v_0 &= 0 \label{v0opt}\\[.05in]
|u_0| &= \sqrt{(1-f_E)\kappa_S} \label{u0opt}\\[.05in]
\alpha &= 0 \label{alphaOpt} \\[.05in]
{\bf v}^\dagger{\bf v} &= f_E\kappa_SN_S \label{v0normOpt}\\[.05in]
{\bf u}^\dagger{\bf u} & =  f_E\kappa_SN_S + 1 - (1-f_E)\kappa_S\label{u0normOpt}\\[.05in]
{\bf v}^\dagger{\bf u} &= 0 \label{vuOpt}.
\end{align}
Her Holevo information rate for this optimum frequency-domain collective Gaussian attack obeys
\begin{eqnarray}
\lefteqn{\chi_{EB} \le \chi_{EB}^{\rm UB} = } \nonumber \\[.05in]
&& R\min\!\left[M\!\left(\kappa_SN_S\{f_E[1/\ln(2) - \log_2(f_E\kappa_SN_S)]\right.\right. + \nonumber \\[.05in]
&& \left.\left.(1-f_E)N_S\log_2(1+1/N_S)\}\right),1\right],
\label{upperbound6}
\end{eqnarray}
This result omits the $O(\kappa_S^{3/2})$ and $O(N_B^{-1/2})$ terms in (\ref{upperbound5}), so it is important to note that:  (1) in computing the paper's secret-key rate results we used the \emph{exact} form from (\ref{upperbound3}) with the attack parameters from Eqs.~(\ref{v0opt})--(\ref{vuOpt}); and (2) numerically maximizing the right-hand side of (\ref{upperbound4}) over Eve's attack parameters for the path lengths considered in the paper yielded $\delta = \gamma_v = \pi/2$ \cite{Zhuang2016}.  

\subsection{Physical Realization of Eve's Optimum Frequency-Domain Collective Attack}
\label{Eve_attack}
At this juncture it is instructive to exhibit a physical implementation for Eve's optimum frequency-domain collective attack, namely her Fig.~\ref{attack} Gaussian attack with attack parameters given by Eqs.~(\ref{v0opt})--(\ref{vuOpt}).  That attack can be realized with Eve's using only two ancilla and choosing $u_1 = \sqrt{f_E\kappa_SN_S + 1 - (1-f_E)\kappa_S}$, $v_1 = 0$, $u_2 = 0$, and $v_2 = \sqrt{f_E\kappa_SN_S}$.  Then, because Alice and Bob must do phase tracking---FL-QKD is an interferometric protocol---no loss of generality ensues from setting $u_0 = \sqrt{(1-f_E)\kappa_S}$.  With these parameter values, Eve's optimum frequency-domain collective Gaussian attack becomes the SPDC beam-splitter attack, shown in Fig.~\ref{realization}.  Here, Eve uses an SPDC source identical to Alice's with the exception of its brightness being $N_E = f_E\kappa_SN_S/[1-(1-f_E)\kappa_S]$.  She retains her idler and injects her signal into the Alice-to-Bob channel through a beam splitter with Alice-to-Bob transmissivity $\sqrt{(1-f_E)\kappa_S}$.  Eve then performs a collective measurement on the light she collects from that beam splitter's other output port, her retained idler, and the light she taps from the Bob-to-Alice channel in which she has inserted a beam splitter with Bob-to-Alice transmissivity $\kappa_S$.  To see that this identification is correct, we exhibit its three-mode Bogoliubov transformation,
\begin{eqnarray}
\hat{a}_{S_m}' &=& \sqrt{(1-f_E)\kappa_S}\,\hat{a}_{S_m} \nonumber \\[.05in]
&+& \sqrt{f_E\kappa_SN_S + 1 - (1-f_E)\kappa_S}\,\hat{e}_{V_m}^{(1)} \nonumber \\[.05in]
&+& \sqrt{f_E\kappa_SN_S}\,\hat{e}_{V_m}^{(2)\dagger} 
\end{eqnarray}
\begin{eqnarray}
\hat{e}_{I_m}^{(1)} &=& \sqrt{\frac{f_E\kappa_SN_S}{1-(1-f_E)\kappa_S}}\,\hat{e}_{V_m}^{(1)\dagger} \nonumber \\[.05in] &+& \sqrt{\frac{f_E\kappa_SN_S + 1 - (1-f_E)\kappa_S}{1-(1-f_E)\kappa_S}}\,\hat{e}_{V_m}^{(2)}
\end{eqnarray}
\begin{eqnarray}
\hat{e}_{I_m}^{(2)} & =& \sqrt{1-(1-f_E)\kappa_S}\,\hat{a}_{S_m} \nonumber \\[.05in]
&+& \sqrt{\frac{(1-f_E)\kappa_S(f_E\kappa_SN_S + 1 - (1-f_E)\kappa_S)}{1-(1-f_E)\kappa_S}}\,\hat{e}_{V_m}^{(1)} \nonumber \\[.05in] &+& \sqrt{\frac{(1-f_E)\kappa_S(f_E\kappa_SN_S)}{1-(1-f_E)\kappa_S}}\,\hat{e}_{V_m}^{(2)\dagger}. 
\end{eqnarray}
and recognize $\hat{a}_{S_m}'$ and $\hat{e}_{I_m}^{(2)}$ as the beam splitter outputs in Fig.~\ref{realization} and $\hat{e}_{I_m}^{(1)}$ as Eve's retained idler.

In the paper, we not only report our upper bound on the Holevo information rate for Eve's optimum frequency-domain collective Gaussian attack, as realized by the SPDC beam-splitter arrangement, but also upper bounds on her Holevo information rates for her collective passive and collective active attacks with that arrangement.  The upper bound on the Holevo information rate of Eve's collective passive attack is trivially obtained from the development presented earlier in this section:  her optimum collective frequency-domain Gaussian attack becomes her collective passive attack when $f_E = 0$.  Eve's optimum collective active attack is realized, in the Fig.~\ref{realization} setup, by her only making a collective measurement on her retained idler and the light she taps from the Bob-to-Alice channel.  That rate bound, which can be derived by a procedure similar to what we have just presented, is as follows:
\begin{equation}
\chi_{EB}^{\rm UBact} = R\min\!\left\{M\left[S_G({\boldsymbol \Lambda}_{IB}) -\sum_{b=0}^1S_G({\boldsymbol \Lambda}_{IB}^{(b)})/2\right],1\right\},
\end{equation}
where 
\begin{equation}
{\boldsymbol \Lambda}^{(b)}_{IB} = \frac{1}{4}\!\left[\begin{array}{cc}
{\bf A}_E & {\bf C}_{IB}^{\rm act(b)} \\[.05in]
{\bf C}_{IB}^{\rm act(b)} & {\bf A}_B
\end{array}\right],
\end{equation}
with ${\bf A}_E = (2N_E+1){\bf I}_2$, ${\bf A}_B = [2(G_BN_S+N_B)+1]{\bf I}_2$, and
\begin{equation}
{\bf C}_{IB}^{\rm act(b)} = \left[\begin{array}{cc}
(-1)^bC_{IB}^{\rm act} & 0 \\[.05in]
0 & (-1)^{b+1}C_{IB}^{\rm act} \end{array}\right],
\end{equation}
with $C_{IB}^{\rm act} = 2\sqrt{G_B(1-f_E\kappa_S)N_E(N_E+1)}$, is the conditional Wigner covariance matrix of the $\{\hat{e}_{I_m}^{(1)},\hat{a}_{B_m}\}$ mode pair given Bob's bit value.  That mode pair's unconditional Wigner covariance matrix is then
\begin{equation}
{\boldsymbol \Lambda}_{IB} = \sum_{b=0}^1{\boldsymbol \Lambda}_{IB}^{(b)}/2.
\end{equation}
As before, the von Neumann entropies in this bound can be found in terms of thermal-state von Neumann entropies via symplectic diagonalization of the Wigner covariances.  

\section{Alice's Error Probabilities and Alice and Bob's Shannon Information Rates}

Because $M\ge 200$ for all the performance evaluations presented in the paper, we can use the Central Limit Theorem to justify the following Gaussian-approximation formula for Alice's error probability \cite{Zhang2013} when Bob's bit value is equally likely to be 0 or 1 and Eve mounts her optimum frequency-domain collective Gaussian attack using the Fig.~\ref{realization} setup:
\begin{equation}
\Pr(e)_{\rm Alice}^{\rm hom} = Q\!\left(\frac{\mu_0 - \mu_1}{\sigma_0 + \sigma_1}\right),
\end{equation}
where   
\begin{equation}
Q(x) = \int_x^\infty\!{\rm d}t\,\frac{e^{-t^2/2}}{\sqrt{2\pi}}.
\end{equation}
Here, $\mu_b$ and $\sigma_b$ are the conditional mean and conditional standard deviation of the $\hat{N}_{\rm hom}$ measurement given the value of Bob's message bit, $b$.  Once Alice's error probability is found, Alice and Bob's Shannon-information rate follows immediately from
\begin{align}
I_{AB} &= R\!\left[1+\Pr(e)_{\rm Alice}^{\rm hom}\log_2(\Pr(e)_{\rm Alice}^{\rm hom}) \right.\nonumber \\[.05in]
&+ \left.(1-\Pr(e)_{\rm Alice}^{\rm hom})\log_2(1-\Pr(e)_{\rm Alice}^{\rm hom})\right],
\end{align}
hence all that remains is to determine the conditional means and standard deviations needed to instantiate our error-probability formula.  

The conditional moments we require are easily calculated from the Fig.~\ref{realization} setup and its associated state characterizations, so we will merely present the results.  We have that
\begin{equation}
\mu_b = 2(-1)^bM\eta\kappa_S \sqrt{G_BN'_{\rm ASE}N_{\rm LO}}, 
\end{equation}
and
\begin{equation}
\sigma_b  = \sqrt{M\{\eta N_1 + 2\eta^2[N^{\rm Alice}_RN_{\rm LO} + \kappa^2_SG_BN'_{\rm ASE}N_{\rm LO}]\}}, 
\end{equation}
where $N'_{\rm ASE} = (1-\kappa_B)(1-f_E)(1-\kappa_A)(1-\kappa_C)N_{\rm ASE}$, $N_1 = N^{\rm Alice}_R + N_{\rm LO}$, 
$N^{\rm Alice}_R = \kappa_SG_B(1-\kappa_B)\kappa_SN_S+ \kappa_SN_B$, and perfect reference storage has been assumed \cite{footnote8}.  At this point we can obtain the asymptotic ($N_B \gg 1, N_{\rm LO} \gg 1$) form of $\Pr(e)^{\rm hom}_{\rm Alice}$ that was used for illustrative purposes in the paper, albeit not in the performance-evaluation figures.  In this asymptotic regime we have that
\begin{equation}
\sigma_b \rightarrow \sqrt{2M\eta^2\kappa_SN_BN_{\rm LO}},
\end{equation}
whence
\begin{equation}
\Pr(e)^{\rm hom}_{\rm Alice} \rightarrow 
 Q\!\left(\sqrt{2M\kappa_SG_BN'_{\rm ASE}/N_B}\right).
\end{equation}
Neglecting the small amount of SPDC light that Alice sent to Bob, we can replace $(1-\kappa_A)(1-\kappa_C)N_{\rm ASE}$ with $N_S$.  Using $M = TW = W/R$, and replacing $(1-\kappa_B)$ with 1 because Bob's channel monitor will withdraw only a small amount of the light he receives from Alice, we then get
\begin{align}
\Pr(e)^{\rm hom}_{\rm Alice} &\rightarrow Q\!\left(\sqrt{2M\kappa_SG_B(1-f_E)N_S/N_B}\right) 
\nonumber \\[.05in]
&\le \exp(-WG_B(1-f_E)N_S/RN_B)/2,
\end{align}
in the $N_B \gg 1$, $N_{\rm LO} \gg 1$ regime, where we have used the well-known bound $Q(x) \le \exp(-x^2/2)/2$.  In the paper, this expression was quoted for ideal equipment, which presumes unity homodyne efficiency ($\eta = 1$).  The derivation we have just given verifies that in this asymptotic regime $\Pr(e)^{\rm hom}_{\rm Alice}$ is not sensitive to the homodyne efficiency.  Thus the $\eta = 0.9$ homodyne efficiency assumed in the paper is \emph{not} a critical value.

We have now obtained upper bounds on the Holevo information rates of Eve's optimum frequency-domain collective attack, her collective passive attack, and her collective active attack, all of which are realizable with the beam-splitter arrangement shown in Fig.~\ref{realization}.  In the paper we plot upper bounds for these attacks' Holevo informations in bits per mode, rather than bits per second.  The bits per mode bounds are trivially obtained by dividing the bits per second bounds by the illumination bandwidth $W$, which specifies the number of modes per second that are being employed on the Alice-to-Bob and Bob-to-Alice channels.  

\section{Channel monitoring for general states}
Alice and Bob measure the singles rates at their channel monitors, i.e., $S_I$ for Alice's idler beam, $S_A$ for Alice's tap on her transmitted beam, and $S_B$ for Bob's tap on his received beam.  They also measure $C_{IA}$ and $\widetilde{C}_{IA}$, the time-aligned and time-shifted coincidence rates between Alice's idler and the tap on her transmitted beam, and $C_{IB}$ and $\widetilde{C}_{IB}$, the time-aligned and time-shifted coincidence rates between Alice's idler and Bob's tap on his received beam, in both cases after accounting for the relevant propagation delays as described below.   Their monitors will be assumed to have detectors with quantum efficiencies $\eta_I, \eta_A$ and $\eta_B$, respectively, and identical jitter-limited coincidence-gate durations, $T_g \sim 100$\,ps.  When the average number of photons illuminating each monitor in a gate time is much smaller than one---as will be the case for our performance evaluation---the average values of the preceding rates can be taken to be \cite{Shapiro2009b}
\begin{equation}
S_K  = \frac{\eta_K}{T_R}\int_{-T_R/2}^{T_R/2}\!{\rm d}t\,\langle\hat{E}^{{\rm mon}\dagger}_K(t)\hat{E}^{\rm mon}_K(t)\rangle, 
\end{equation}
for $K=I,A,B$, and 
\begin{align}
C_{IK} &= \frac{\eta_I\eta_K}{T_R}\int_{-T_R/2}^{T_R/2}\!{\rm d}t\int_{t-T_g/2}^{t+T_g/2}\!{\rm d}u\, \nonumber \\[.05in] & \times \langle \hat{E}^{{\rm mon}\dagger}_I(t)\hat{E}^{\rm mon}_I(t)\hat{E}^{{\rm mon}\dagger}_K(u)\hat{E}^{\rm mon}_K(u)\rangle,\label{timealigned}\\[.05in]
\widetilde{C}_{IK} &= \frac{\eta_I\eta_K}{T_R}\int_{-T_R/2}^{T_R/2}\!{\rm d}t\int_{t+T_s-T_g/2}^{t+T_s+T_g/2}\!{\rm d}u\, \nonumber \\[.05in] & \times \langle \hat{E}^{{\rm mon}\dagger}_I(t)\hat{E}^{\rm mon}_I(t)\hat{E}^{{\rm mon}\dagger}_K(u)\hat{E}^{\rm mon}_K(u)\rangle,\label{timeshifted}
\end{align}
for $K=A,B$, where $\hat{E}^{\rm mon}_K(t)$, for $K=I,A,B$, are the positive-frequency, $\sqrt{\mbox{photons/s}}$-units field operators entering Alice's idler and transmitter tap monitors and Bob's monitor, respectively.  Here, the time-origins for the $\{\hat{E}^{\rm mon}_K(t)\}$ have been chosen to ensure that true coincidences \emph{and} accidental coincidences will be counted in the time-aligned coincidences $C_{IK}$, but \emph{only} accidental coincidences will be counted in the time-shifted coincidences $\widetilde{C}_{IK}$.  The latter condition is ensured by taking the time shift $T_s$ to satisfy $WT_s \gg 1$, $T_s \gg T_g$, and $T_s \ll T_R$, where $W$ is Alice's source bandwidth and $t\in [-T_R/2,T_R/2]$ is the duration of the FL-QKD protocol's quantum communication.  In practice, $T_s\sim 10\,$ns will suffice for $W = 2\,$THz and $T_g = 100\,$ps.  

If we assume that Eve mounts a collective frequency-domain Gaussian attack, then all of the fields appearing in our singles and coincidence rates are in a zero-mean, jointly-Gaussian state and we can evaluate these rates by means of Gaussian moment factoring \cite{Shapiro1994}.  However, because we seek security against the general frequency-domain collective attack, we will show that Alice and Bob's channel monitors can determine Eve's intrusion parameter, $f_E$, even when her attack in \emph{not} Gaussian.  Toward that end it is convenient to introduce Fourier-series decompositions for the field operators $\{\,\hat{E}^{\rm mon}_K(t) : K = I,A,B\,\}$ over the entire duration of FL-QKD's quantum communication, viz.,
\begin{align}
\hat{E}^{\rm mon}_{I}(t) &=& \frac{e^{-i\omega_I t}}{\sqrt{T_R}}\sum_{m=-WT_R/2}^{WT_R/2}\hat{a}^{\rm mon}_{I_m}e^{-i2\pi mt/T_R},\\[.05in]
\hat{E}^{\rm mon}_{K}(t) &=& \frac{e^{-i\omega_St}}{\sqrt{T_R}}\sum_{m=-WT_R/2}^{WT_R/2}\hat{a}^{\rm mon}_{K_m}e^{i2\pi mt/T_R},
\end{align}
for $K = A,B$, where $\omega_S$ and $\omega_I$ are the center frequencies of Alice's signal and idler beams and we have limited the series to Alice's source bandwidth, i.e., to the frequency modes that are in non-vacuum states.     
The behaviors of the modes appearing in these Fourier series can be gotten from App.~A by presuming that the Fourier expansions in that appendix were made on the $[-T_R/2,T_R/2]$ interval and making the following identifications:
\begin{align}
\hat{a}^{\rm mon}_{I_m} &= \hat{a}_{I_m}^{\rm SPDC} 
\label{aIrelation}\\[.05in]
\hat{a}^{\rm mon}_{A_m} &= \sqrt{\kappa_A}\,\hat{a}_{A_m} - \sqrt{1-\kappa_A}\,\hat{v}_{A_m}
\label{aArelation}\\[.05in]
\hat{a}^{\rm mon}_{B_m} &= \sqrt{\kappa_B}\,\hat{a}'_{S_m} - \sqrt{1-\kappa_B}\,\hat{v}_{B_m}.
\label{aBrelation}
\end{align}
Note that Eve's mounting a frequency-domain collective attack makes the mode triples $\{\,(\hat{a}^{\rm mon}_{I_m},\hat{a}^{\rm mon}_{A_m},\hat{a}^{\rm mon}_{B_m}) : -WT_R/2 \le m \le WT_R/2\,\}$ independent and identically distributed with the $\{\hat{a}_{I_m}\}$ modes being in zero-mean states.  

For the singles rates we find that
\begin{eqnarray}
S_K  &=& \frac{\eta_K}{T_R}\sum_{n=-WT_R/2}^{WT_R/2}\sum_{m=-WT_R/2}^{WT_R/2}\langle \hat{a}^{{\rm mon}\dagger}_{K_n}\hat{a}^{\rm mon}_{K_m} \rangle \nonumber \\[.05in]
&\times& \frac{\sin[\pi (n-m)]}{\pi (n-m)} \\[.05in]
&=& \frac{\eta_K}{T_R} \sum_{n=-WT_R/2}^{WT_R/2}\langle \hat{a}^{{\rm mon}\dagger}_{K_n}\hat{a}^{\rm mon}_{K_n} \rangle \\[.05in]
&=& 
\eta_K W\langle \hat{a}^{{\rm mon}\dagger}_{K_n}\hat{a}^{\rm mon}_{K_n} \rangle,
\end{eqnarray}
for $K = I,A,B$.  Using this result in conjunction with Eqs.~(\ref{aIrelation})--(\ref{aBrelation}) then gives us
\begin{align}
S_I &= \eta_IN_{\rm SPDC}W,\\[.05in]
S_A &= \eta_A\kappa_AN_AW,
\label{SA}\\[.05in]
S_B &= \eta_B\kappa_B\kappa_SN_SW.
\label{SB}
\end{align}

Finding the time-aligned and time-shifted coincidence rates is more complicated than what we have just done for the singles rates.  We start from the photon-flux cross-correlation function,
\begin{equation}
R_{IK}(t,u) = \langle \hat{E}^{{\rm mon}\dagger}_I(t)\hat{E}^{\rm mon}_I(t)\hat{E}^{{\rm mon}\dagger}_K(u)\hat{E}^{\rm mon}_K(u)\rangle,
\end{equation}
for $K=A,B$, which, employing the Fourier series given earlier and grouping terms, can be reduced to
\begin{equation}
R_{IK}(t,u)  = \sum_{k=1}^3 R_{IK}^{(k)}(t,u),
\end{equation}
where 
\begin{eqnarray}
R_{IK}^{(1)}(t,u) &=& \frac{1}{T_R^2}\!\left[\sum_{n, m}\langle \hat{a}^{{\rm mon}\dagger}_{I_n}\hat{a}^{{\rm mon}\dagger}_{K_n}\rangle \langle\hat{a}^{\rm mon}_{I_m}\hat{a}^{\rm mon}_{K_m}\rangle\right. \nonumber \\[.05in]
&\times&  \left.e^{i2\pi(n-m)(t-u)/T_R}\right],
\end{eqnarray}
\begin{equation}
R_{IK}^{(2)} = \frac{1}{T_R^2}\!\left[\sum_{n,m}\langle \hat{a}^{{\rm mon}\dagger}_{I_n}\hat{a}^{\rm mon}_{I_n}\rangle \langle \hat{a}^{{\rm mon}\dagger}_{K_m}\hat{a}^{\rm mon}_{K_m}\rangle \right],
\end{equation}
and
\begin{eqnarray}
\lefteqn{R_{IK}^{(3)}(t,u) =  }\nonumber \\[.05in]
&& \frac{1}{T_R^2}\!\left\{\sum_n\!\left[\langle \hat{a}^{{\rm mon}\dagger}_{I_n}\hat{a}^{{\rm mon}\dagger}_{K_n}\hat{a}^{\rm mon}_{I_n}\hat{a}^{\rm mon}_{K_n}\rangle - |\langle \hat{a}^{\rm mon}_{I_n}\hat{a}^{\rm mon}_{K_n}\rangle|^2\right.\right. \nonumber \\[.05in]
&-&\left. \left.\langle \hat{a}^{{\rm mon}\dagger}_{I_n}\hat{a}^{\rm mon}_{I_n}\rangle \langle \hat{a}^{{\rm mon}\dagger}_{K_n}\hat{a}^{\rm mon}_{K_n}\rangle\right]\right\},
\end{eqnarray}
because of the independence of the mode triples and the zero-mean nature of the $\{\hat{a}^{\rm mon}_{I_m}\}$ modes, with all indices are summed from $-WT_R/2$ to $WT_R/2$.

The time-independence of $R_{IK}^{(2)}(t,u)$ and $R_{IK}^{(3)}(t,u)$ implies that these terms will not contribute to $C_{IK}-\widetilde{C}_{IK}$.  Moreover the independence and identical distribution of the mode pairs $\{\hat{a}^{\rm mon}_{I_m},\hat{a}^{\rm mon}_{A_m}\,\hat{a}^{\rm mon}_{B_m}\}$ makes $R_{IK}^{(1)}(t,u)$ vanish when $|t-u| \gg 1/W$.  Hence we find that
\begin{eqnarray}
C_{IK} - \widetilde{C}_{IK} &=&  \frac{\eta_I\eta_K}{T_R}|\langle \hat{a}^{\rm mon}_{I_m}\hat{a}^{\rm mon}_{K_m}\rangle|^2 \nonumber \\[.05in] 
&\times& \sum_{n, m} \frac{T_g}{T_R}\frac{\sin[\pi (n-m)T_g/T_R]}{\pi (n-m)T_g/T_R}.
\label{truecoinc}
\end{eqnarray}  

In the main text we claimed that Alice and Bob's channel monitors will enable them to measure Eve's intrusion parameter,
\begin{equation}
f_E \equiv 1- \frac{|\langle\hat{a}'_{S_m}\hat{a}_{I_m}\rangle|^2}{\kappa_S|\langle \hat{a}_{S_m}\hat{a}_{I_m}\rangle|^2},
\label{fEdefn}
\end{equation}
via
\begin{equation}
f_E = 1-\frac{[C_{IB}-\widetilde{C}_{IB}]/S_B}{[C_{IA}-\widetilde{C}_{IA}]/S_A}.
\label{fEmeas}
\end{equation}
Using Eqs.~(\ref{SA}), (\ref{SB}), and (\ref{truecoinc}) we get
\begin{equation}
\frac{[C_{IB}-\widetilde{C}_{IB}]/S_B}{[C_{IA}-\widetilde{C}_{IA}]/S_A} = 
\frac{|\langle\hat{a}^{\rm mon}_{I_m}\hat{a}^{\rm mon}_{B_m}\rangle|^2}{|\langle \hat{a}^{\rm mon}_{I_m}\hat{a}^{\rm mon}_{A_m}\rangle|^2}
\frac{\langle \hat{a}^{{\rm mon}\dagger}_{A_m}\hat{a}^{\rm mon}_{A_m}\rangle}
{\langle \hat{a}^{{\rm mon}\dagger}_{B_m}\hat{a}^{\rm mon}_{B_m}\rangle}.
\end{equation}
From Eqs.~(\ref{aIrelation})--(\ref{aBrelation}) we can reduce this result to 
\begin{equation}
\frac{[C_{IB}-\widetilde{C}_{IB}]/S_B}{[C_{IA}-\widetilde{C}_{IA}]/S_A} = 
\frac{|\langle\hat{a}_{I_m}^{\rm SPDC}\hat{a}'_{S_m}\rangle|^2}{|\langle \hat{a}_{I_m}^{\rm SPDC}\hat{a}_{A_m}\rangle|^2}
\frac{\langle \hat{a}^\dagger_{A_m}\hat{a}_{A_m}\rangle}
{\langle \hat{a}'^\dagger_{S_m}\hat{a}'_{S_m}\rangle}.
\end{equation}
Use of Eqs.~(\ref{SmSA}) and (\ref{mean_photon_constraint}) plus $\langle \hat{a}_{S_m}\hat{a}_{I_m}^{\rm SPDC}\rangle = \sqrt{\kappa'}\langle \hat{a}_{S_m}\hat{a}_{I_m}\rangle$ then yields
\begin{equation}
\frac{[C_{IB}-\widetilde{C}_{IB}]/S_B}{[C_{IA}-\widetilde{C}_{IA}]/S_A} = \frac{|\langle\hat{a}'_{S_m}\hat{a}_{I_m}\rangle|^2}{\kappa_S|\langle \hat{a}_{S_m}\hat{a}_{I_m}\rangle|^2}.
\label{fEagreement}
\end{equation}
Although this result appears to verify the agreement of Eqs.~(\ref{fEdefn}) and (\ref{fEmeas}), there is an issue with that identification.  The modes appearing in Eq.~(\ref{fEdefn}) were obtained from Fourier-series decompositions of the relevant continuous-time field operators on a duration-$1/R$\,s interval, whereas those in Eq.~(\ref{fEagreement}) come from Fourier-series decompositions of those field operators on a duration-$T_R$\,s interval.  Because of the independent, identical distribution of the mode operators, however, their second moments---which are all that appears in Eqs.~(\ref{fEdefn}) and (\ref{fEmeas})---will be the same regardless of whether the Fourier series' time interval has duration $1/R$ or $T_R$.  

\section{Eve's Entanglement-Assisted Capacity}
When Eve mounts a collective active attack, we can regard her use of the SPDC's idler beam she has retained and the modulated, amplified, noisy version of her SPDC's signal beam she collects from her tap on the Bob-to-Alice fiber as an entanglement-assisted communication channel from Bob to her.  Consequently, her collective active attack's Holevo information per mode cannot exceed the single-mode entanglement-assisted capacity for that channel, $C_E$ \cite{Holevo2001,Bennett2002}, because entanglement-assisted capacity is known to be additive.  From \cite{Holevo2001,Bennett2002} we have that
\begin{eqnarray}
C_E &=& g[(1-\kappa_B)[1-(1-f_E)\kappa_S]N_E] \nonumber \\[.05in] &+& g[G_B(1-\kappa_B)[1-(1-f_E)\kappa_S]N_E + N_B] 
\nonumber \\[.05in]
&-&g[(1+(1-\kappa_B)[1-(1-f_E)\kappa_S]N_E)N_B].
\label{Ce}
\end{eqnarray}
We have been somewhat conservative in Eq.~(\ref{Ce}) in that this result assumes that Alice does not inject any light into Bob and that Eve collects all the light that Bob sends on the Bob-to-Alice fiber.  Neither of these assumptions is of great consequence, but they make it easier to obtain the result in Eq.~(\ref{Ce}).  In particular, Alice's injection into Bob acts as noise for Eve's active attack.  Moreover, because Alice's injection into Bob has low brightness, it is dwarfed by the ASE from Bob's amplifier.  Finally, because Fig.~\ref{ppb}(b) plots $C_E$ for a 50-km-long path, Eve is already getting 90\% of the light Bob sends to Alice.  Hence increasing that value to 100\% is not a major change, especially since Bob's amplifier gain is sufficient to overcome return-path loss.

\section{Bounding Eve's information gain from knowing the output of Bob's channel monitor}
Bob sends Alice the times at which his channel monitor has detected photons so that she can use that data to estimate Eve's intrusion parameter.  To do so he uses a tamper-proof classical channel that Eve can monitor.  So far, we have not included the information that Eve could glean from that classical transmission in bounding her Holevo information rate.  Here we will show that the extra information that Eve might gain from knowing those detection times is inconsequential.  

The mean photon-number per bit at Bob's monitor detector is $M\kappa_B\kappa_S N_S\simeq \kappa_B\ll1$, owing to FL-QKD's operating with $M \kappa_SN_S\sim 1$ ($\sim$1 ppb at Bob's terminal), so we will only consider two leading-order possibilities:   no photon is detected (probability of occurrence = $p_0$) or one photon is detected (probability of occurrence = $p_1=1-p_0$).  

Let us use $\chi_{EB\mid n}^{\rm UB}$, for $n=0,1$, to denote an upper bound on Eve's Holevo information rate given that Bob's monitor has detected $n$ photons \emph{and}, if there has been a detection, that Eve knows from which frequency mode it came.  (This frequency-mode knowledge is not available to Eve from her eavesdropping on Bob's classical-channel transmission, so assuming she has this knowledge increases her Holevo information rate.)  Then, averaged over Bob's monitor result, the upper bound on Eve's Holevo information rate for her optimum frequency-domain collective attack is 
\begin{equation}
\bar{\chi}_{EB}^{\rm UB} = p_0\chi_{EB\mid 0}^{\rm UB}+p_1\chi_{EB\mid 1}^{\rm UB}.
\end{equation}
Because all $M$ modes are independent, we have that $\chi_{EB\mid 0}^{\rm UB}=M \chi_0$, where $\chi_0$ is the per-mode upper bound on Eve's Holevo information rate when Bob's monitor failed to detect a photon \cite{footnote9}. When Bob's monitor does detect a photon, and Eve knows which frequency mode has lost a photon to that detection, the upper bound on her conditional Holevo information rate will be 
\begin{equation}
\chi_{EB\mid 1}^{\rm UB}=(M-1)\chi_0+\chi_1,
\end{equation}
where $\chi_1$ is the per-mode upper bound when Bob's monitor detected a photon in that mode.  We now have that
\begin{equation}
\bar{\chi}_{EB}^{\rm UB} = M \chi_0+p_1(\chi_1-\chi_0),
\end{equation}
which we need to compare to our upper bound from App.~C, which neglected any information Eve might gain from learning the times at which Bob's channel monitor made photon detections.

For $\chi_{EB}^{\rm UB}$ being the App.~C upper bound we will use $\chi \equiv \chi_{EB}^{UB}/M$,to denote its per-mode contribution. We now have that
\begin{equation}
\frac{\bar{\chi}_{EB}^{\rm UB}}{\chi_{EB}^{\rm UB}} =  \frac{\chi_0}{\chi}+p_1\frac{(\chi_1-\chi_0)}{M\chi}.
\label{chibar1}
\end{equation}  
Figure~4(a) shows that Bob will receive $\sim$1\,ppb for one-way path lengths less than 200\,km, and our secret-key rate calculations assume that Bob's monitor taps 1\% of that light.  Together these conditions imply that $p_1 \approx 0.01$.  Figure~4(a) also implies that $M\chi \approx 0.8$ for a 50\,km one-way path length.  So, taking the \emph{very} conservative upper limit of unity for $\chi_1-\chi_0$, we have that the second term on the right in Eq.~(\ref{chibar1}) is at most 0.013.  Thus it only remains for us to address the first term on the right in that equation.  We will do so within the App.~C.2 framework, i.e., for Eve' frequency-domain collective Gaussian attack.

Eve gains her information from measuring the mode triples $\{\hat{e}_{I_m}^{(1)}$, $\hat{e}_{I_m}^{(2)},\hat{a}_{B_m}\}$. To assess the impact of Eve's having Bob's channel-monitor data, we focus our attention on what that data implies about conditional state of the $\{\hat{a}''_{S_m}\}$ modes, viz., the modes that enter Bob's BPSK modulator and, after modulation and subsequent amplification, become the $\{\hat{a}_{B_m}\}$ modes.  Moreover, to do so we will presume that the $\{\hat{a}'_{S_m}\}$ modes that arrive at Bob's terminal are in independent, identically-distributed thermal states with average photon number $\kappa_SN_S$, as is the case in Eve's optimum frequency-domain Gaussian collective attack.  Using the beam-splitter relation that converts these modes and the vacuum-state $\{\hat{v}_{B_m}\}$ modes into the $\{\hat{a}^{\rm mon}_{B_m},\hat{a}''_{S_m}\}$ mode pairs, we find that those mode pairs are in independent, identically-distributed Gaussian states whose coherent-state decomposition is 
\begin{eqnarray}
\hat{\rho}_{\hat{a}^{\rm mon}_{B_m},\hat{a}''_{S_m}}&=&\int \frac{{\rm d}^2\alpha}{\pi \kappa_SN_S} e^{-|\alpha|^2/\kappa_SN_S} \ket{\sqrt{\kappa_B}\,\alpha}\!{}_B{}_B\!\bra{\sqrt{\kappa_B}\,\alpha}\nonumber\\[.05in] &\otimes& \ket{\sqrt{1-\kappa_B}\,\alpha}\!{}_S{}_S\!\bra{\sqrt{1-\kappa_B}\,\alpha}.
\end{eqnarray}
Given that Bob's monitor did not detect a photon, the $\{\hat{a}''_{S_m}\}$ modes are still independent and identically distributed, with conditional density operator
\begin{eqnarray}
\hat{\rho}_{\hat{a}_{S_m}''\mid 0}=\frac{{}_B\!\bra{0} \hat{\rho}_{\hat{a}^{\rm mon}_{B_m},\hat{a}_{S_m}''}\ket{0}\!{}_B}{{\rm Tr}\!\left({}_B\!\bra{0} \hat{\rho}_{\hat{a}^{\rm mon}_{B_m},\hat{a}_{S_m}''}\ket{0}\!{}_B\right)}.
\end{eqnarray}
After some algebra, we have the $\hat{\rho}_{\hat{a}_{S_m}''\mid 0}$ is a thermal state whose mean photon number,  
$(1-\kappa_B)\kappa_SN_S/(1+\kappa_B\kappa_SN_S)$, is less than that mode's unconditional photon number, $(1-\kappa_B)\kappa_SN_S$. Thus we conclude conditioning on Bob getting no count, the mean photon number in the return mode decreases, but the quantum state is still Gaussian. Similar results hold for Eve's $\{\hat{e}_{I_m}^{(1)}$, $\hat{e}_{I_m}^{(2)}\}$ modes, and we conclude that $\chi_0/\chi < 1$, hence $\bar{\chi}_{EB}^{\rm UB}/\chi_{EB}^{\rm UB} < 1.013$ at 50\,km one-way path length.

\end{appendix}


\begin{thebibliography}{2}

\bibitem{Shannon1949}C. E. Shannon, Bell Syst. Tech. J. {\bf 28,} 656-715 (1949).

\bibitem{Bennett1984}C. H. Bennett and G. Brassard, ``Quantum cryptography:  Public key distribution and coin tossing,''Proc. IEEE International Conf. on Computers, Systems, and Signal Processing, Bangalaore, pp.~175--179 (IEEE, New York, 1984).

\bibitem{Ekert1991}A. K. Ekert, Phys. Rev. Lett. {\bf 67,} 661--663 (1991).

\bibitem{Gisin2002}N. Gisin, G. G. Ribordy, W. Tittel, and H. Zbinden, Rev. Mod. Phys. {\bf 74,} 145--195 (2002).

\bibitem{Grosshans2002}F. Grosshans and P. Grangier,  Phys. Rev. Lett. {\bf 88,} 057902 (2002).

\bibitem{Lucamarini2013}M. Lucamarini, K. A. Patel, J. F. Dynes, B. Fr\"{o}lich, A. W. Sharpe, A. R. Dixon, Z. L. Yuan, R. V. Penty, and A. J. Shields, Opt. Express {\bf 21,} 24550--24565 (2013).

\bibitem{Huang2015}D. Huang, D. Lin, C. Wang, W. Liu, S. Fang, J. Peng, P. Huang, and G. Zeng, Opt. Express {\bf 23,} 17511--17519 (2015).

\bibitem{footnote1} A passive attack therefore has $f_E = 0$ while an active attack has $f_E >0$.  

\bibitem{Navascues2005}M. Navascu\'{e}s and A. Ac\'{i}n, Phys. Rev. Lett. {\bf 94,} 020505 (2005).

\bibitem{Zhang2014}Z. Zhang, J. Mower, D. Englund, F. N. C. Wong, and J. H. Shapiro, Phys. Rev. Lett. {\bf 112,} 120506 (2014).

\bibitem{Pirandola2015a}S. Pirandola, C. Ottaviani, G. Spedalieri, C. Weedbrook. S. L. Braunstein, S. Lloyd, T. Gehring, C. S. Jacobsen, and U. L. Andersen, Nature Photon. {\bf 9,} 397--402 (2015).

\bibitem{AppD} See App.~D for details.

\bibitem{Shapiro2009}J. H. Shapiro, Phys. Rev. A {\bf 80,} 022320 (2009).

\bibitem{Zhang2013}Z. Zhang, M. Tengner, T. Zhong, F. N. C. Wong, and J. H. Shapiro,  Phys. Rev. Lett. {\bf 111,} 010501 (2013).

\bibitem{Pirandola2008}S. Pirandola and S. Lloyd, Phys. Rev. A {\bf 78,} 012331 (2008).

\bibitem{AppC} See App.~C for details.

\bibitem{Richardson2001}T. J. Richardson, A. Shokrollah, and R. L. Urbanke,  IEEE Trans. Inform. Theory {\bf 47,} 619--637 (2001).

\bibitem{footnote2}Bob's receiving $\sim$1 ppb keeps $\Pr(e)_{\rm Alice}^{\rm hom} = 0.1$ when his amplifier's gain is sufficient to make return-path loss inconsequential.  The increase in Bob's received ppb seen in Fig.~4(a) as the path length decreases from 50\,km is due to our $R\le 10\,$Gbps constraint, which leads to $\Pr(e)_{\rm Alice}^{\rm hom}$ appreciably lower than 0.1 at such short distances when Alice and Bob optimize their secret-key rate.  On the other hand, the increase in Bob's received ppb as the path length increases beyond 150\,km occurs because $G_B = 10^4$ is becoming insufficient to overcome return-path loss at these distances, so that Bob must receive more ppb to ensure $\Pr(e)_{\rm Alice}^{\rm hom} \le 0.1$.  

\bibitem{AppC2F} See Apps.~C.2 and F for  details.

\bibitem{AppC2} See App.~C.2 for details.

\bibitem{footnote3}It follows from Eve's passive-attack Holevo information rate versus Alice's signal brightness, shown in Fig.~4(b), and Alice's source brightness versus path length, shown in Fig.~3(a), that Alice and Bob can realize a 2\,Gbps secret-key rate over a 50-km-long fiber link with complete immunity to the undetectable passive-eavesdropping attack.  In this regard we note that because a passive eavesdropper does \emph{not} interact with the light going to Bob, the collective passive-eavesdropping attack is its most powerful form, i.e., there is no coherent passive-eavesdropping attack.

\bibitem{Leverrier2009}A. Leverrier and P. Grangier, Phys. Rev. Lett. {\bf 102,} 180504 (2009).

\bibitem{Leverrier2011a}A. Leverrier and P. Grangier, Phys. Rev. A {\bf 83,} 042312 (2011).

\bibitem{Leverrier2011b}A. Leverrier and P. Grangier, Phys. Rev. Lett. {\bf 106,} 259902 (2011).

\bibitem{Pirandola2015b}S. Pirandola, R. Laurenza, C. Ottaviani, and L. Banchi, arXiv:1510.08863 [quant-ph].

\bibitem{footnote4}Pirandola \emph{et al.} \cite{Pirandola2015b} quote this limit in bits per channel use, but their channel uses are implicitly single mode.  

\bibitem{Mower2013}J. Mower, Z. Zhang, P. Desjardins, C. Lee, J. H. Shapiro, and D. Englund, Phys. Rev. A {\bf 87,} 062322 (2013).

\bibitem{footnote5}The strong ASE contributed by Bob's amplifier to the light he returns to Alice provides the dominant noise in Alice's homodyne receiver, see App.~D.

\bibitem{footnote6}This threshold value minimizes Alice's error probability.

\bibitem{footnote7}If $\kappa_I$ is so small that amplifiers with gain $G_R = 1/\kappa_I$ are unavailable, then the fiber spool can be divided into a series connection of shorter-length spools---that together provide the required overall delay--with a loss-compensating amplifier employed at the input to each of them.  Such an arrangement can achieve the same goal of near-perfect reference storage.

\bibitem{quantum_data_processing}R. Ahlswede and P. Lober, IEEE Trans. Inform. Theory {\bf 	47,}, 474-478 (2001).	

\bibitem{Wolf_2006}M. M. Wolf, G. Giedke, and J. I. Cirac, Phys. Rev. Lett. {\bf 96,} 080502 (2006).

\bibitem{Weedbrook2012}C. Weedbrook, S. Pirandola, R. Garc\'{i}a-Patr\'{o}n, N. J. Cerf, T. C. Ralph, J. H. Shapiro, and S. Lloyd, Rev. Mod. Phys. {\bf 84,} 621--669 (2012).

\bibitem{Zhuang2016}Q. Zhuang, Z. Zhang, and J. H. Shapiro, in preparation.

\bibitem{footnote8}Because $\sigma_0 = \sigma_1$, Alice's setting her homodyne receiver's decision threshold to $(\mu_0 + \mu_1)/2 = 0$ minimizes $\Pr(e)_{\rm Alice}^{\rm hom}$, and does so with equal false-alarm and miss probabilities.   

\bibitem{Shapiro2009b}J. H. Shapiro,  IEEE J. Sel. Top. Quantum Electron. {\bf 15,} 1547 (2009).

\bibitem{Shapiro1994}J. H. Shapiro and K.-X. Sun, J. Opt. Soc. Am. B {\bf 11,} 1130 (1994).

\bibitem{Holevo2001}A. S. Holevo and R. F. Werner, Phys. Rev. A {\bf 63,} 032312 (2001).

\bibitem{Bennett2002}C. H. Bennett, P. W. Shor, J. A. Smolin, and A. V. Thapliyal, IEEE Trans. Inform. Theory {\bf 	48,}, 2637-2655 (2002).	
\bibitem{footnote9}Here, and in the rest of App.~G, we shall ignore the upper limit of $R$\,bps on Eve's Holevo information rate.  

\end{thebibliography}
\end{document}